\documentclass[lettersize,journal]{IEEEtran}
\usepackage{amsmath,amsfonts}
\usepackage{amssymb,amsthm}
\usepackage{algorithmic}
\usepackage{algorithm}
\usepackage{array}
\usepackage{color}
\usepackage[caption=false,font=footnotesize,labelfont=sf,textfont=sf]{subfig}
\usepackage{textcomp}
\usepackage{stfloats}
\usepackage{url}
\usepackage{verbatim}
\usepackage{graphicx}
\usepackage{cite}
\usepackage{booktabs}
\usepackage{float}                  
\usepackage{subfig}                 
\usepackage{overpic}
\usepackage{tabularx}
\usepackage{makecell}
\usepackage{geometry}
\usepackage{picins}
\newtheorem{theorem}{Theorem}
\newtheorem{lemma}{Lemma}
\newtheorem{remark}{Remark}

\newcommand{\lmref}[1]{Lemma \ref{#1}}
\newcommand{\thref}[1]{Theorem \ref{#1}}

\newcommand{\figref}[1]{Fig.~\ref{#1}}
\newcommand{\tabref}[1]{Table \ref{#1}}
\newcommand{\alref}[1]{Algorithm \ref{#1}}
\newcommand{\appref}[1]{Appendix \ref{#1}}
\newcommand{\secref}[1]{Section \ref{#1}}

\newcommand{\diag}[1]{\mathrm{diag}\left(#1\right)}
\newcommand{\logdet}[1]{\log\det\left(#1\right)}

\newcommand{\argmin}[1]{\mathop{\arg\min}\limits_{#1}}

\newcommand{\cA}{\mathcal{A}}
\newcommand{\cB}{\mathcal{B}}
\newcommand{\cC}{\mathcal{C}}

\newcommand{\cE}{\mathcal{E}}
\newcommand{\cF}{\mathcal{F}}

\newcommand{\cM}{\mathcal{M}}
\newcommand{\cN}{\mathcal{N}}

\newcommand{\cR}{\mathcal{R}}
\newcommand{\cS}{\mathcal{S}}
\newcommand{\cT}{\mathcal{T}}
\newcommand{\cU}{\mathcal{U}}

\newcommand{\ba}{\mathbf{a}}

\newcommand{\bs}{\mathbf{s}}

\newcommand{\bw}{\mathbf{w}}

\newcommand{\by}{\mathbf{y}}
\newcommand{\bz}{\mathbf{z}}

\newcommand{\bA}{\mathbf{A}}
\newcommand{\bB}{\mathbf{B}}
\newcommand{\bC}{\mathbf{C}}
\newcommand{\bD}{\mathbf{D}}

\newcommand{\bH}{\mathbf{H}}
\newcommand{\bI}{\mathbf{I}}

\newcommand{\bM}{\mathbf{M}}

\newcommand{\bP}{\mathbf{P}}
\newcommand{\bQ}{\mathbf{Q}}
\newcommand{\bR}{\mathbf{R}}

\newcommand{\bU}{\mathbf{U}}
\newcommand{\bV}{\mathbf{V}}
\newcommand{\bW}{\mathbf{W}}
\newcommand{\bX}{\mathbf{X}}
\newcommand{\bY}{\mathbf{Y}}
\newcommand{\bZ}{\mathbf{Z}}

\newcommand{\bbC}{\mathbb{C}}
\newcommand{\bbR}{\mathbb{R}}

\newcommand{\bzero}{\mathbf{0}}

\newcommand{\bPi}{{\boldsymbol\Pi}}

\newcommand{\bGamma}{{\boldsymbol\Gamma}}

\newcommand{\bxi}{{\boldsymbol\xi}}

\newcommand{\bzeta}{{\boldsymbol\zeta}}

\begin{document}
	\thispagestyle{empty}
	\pagestyle{empty}
	\title{Precoder Design for User-Centric Network Massive MIMO with Matrix Manifold Optimization}
	
	\author{Rui~Sun,~\IEEEmembership{Student Member, IEEE}, Li~You,~\IEEEmembership{Senior Member, IEEE}, An-An~Lu,~\IEEEmembership{Member, IEEE}, Chen~Sun,~\IEEEmembership{Member, IEEE}, Xiqi~Gao,~\IEEEmembership{Fellow, IEEE}, and Xiang-Gen~Xia,~\IEEEmembership{Fellow, IEEE}

		\thanks{This work was supported by the Jiangsu Province Basic Research Project under Grant BK20192002, the Jiangsu Province Major Science and Technology Project under Grant SBG2024000080, the Fundamental Research Funds for the Central Universities under Grants 2242022k60007 and 2242023K5003, the National Natural Science Foundation of China under Grants 62322104, 62394294, 62371125  and 62271145, the Key R\&D Plan of Jiangsu Province under Grants BE2022067, BE2022067-2 and BE2022067-5, the Natural Science Foundation of Jiangsu Province under Grant BK20231415, the Natural Science Foundation on Frontier Leading Technology Basic Research Project of Jiangsu under Grant BK20222001, and the Huawei Cooperation Project.	  
			An earlier version of this paper will be presented in part at the IEEE Global Communications Conference (GLOBECOM) 2024.
			\textit{(Corresponding author: Xiqi Gao.)} }

		\thanks{Rui Sun, Li You, An-An Lu, Chen Sun and Xiqi Gao are with the National Mobile Communications Research Laboratory,
			Southeast University, Nanjing 210096, China and are also with Purple Mountain Laboratories, Nanjing 211111, China (e-mail:
			ruisun@seu.edu.cn; lyou@seu.edu.cn;  aalu@seu.edu.cn; sunchen@seu.edu.cn;  xqgao@seu.edu.cn).}
		\thanks{Xiang-Gen Xia is with the Department of Electrical and Computer Engineering, University of Delaware, Newark, DE 19716
			USA (e-mail: xxia@ee.udel.edu).}
	}
	
	\maketitle
	\begin{abstract}
		In this paper, we investigate the precoder design for user-centric network (UCN) massive multiple-input multiple-output (mMIMO) downlink with matrix manifold optimization. In UCN mMIMO systems, each user terminal (UT) is served by a subset of base stations (BSs) instead of all the BSs, facilitating the implementation of the system and lowering the dimension of the precoders to be designed. By proving that the precoder set satisfying the per-BS power constraints forms a Riemannian submanifold of a linear product manifold, we transform the constrained precoder design problem in Euclidean space to an unconstrained one on the Riemannian submanifold.  Riemannian ingredients, including orthogonal projection, Riemannian gradient, retraction and vector transport, of the problem on the Riemannian submanifold are further derived, with which the Riemannian conjugate gradient (RCG) design method is proposed for solving the unconstrained problem. The proposed method avoids the inverses of large dimensional matrices, which is beneficial in practice. The complexity analyses show the high computational efficiency of RCG precoder design.  Simulation results demonstrate  the  numerical superiority of the proposed precoder design and the high efficiency of the UCN mMIMO system.
	\end{abstract}
	
	
	\section{Introduction}
	\thispagestyle{empty}	
	With the rapid deployment of the fifth generation (5G)  networks around the world, both industry and academia have embarked on the research of beyond 5G and the sixth generation (6G) communications \cite{6G}.
	Massive multiple-input multiple-output (mMIMO) has been
	one of the most essential technologies in 5G wireless communications and is believed to be one of the key enabling technologies for the terrestrial 6G networks \cite{mMIMO,shiding}, as well as the the non-terrestrial communication systems \cite{lin1,YL2,lin2}. By grouping together antennas at the transmitter and the receiver, respectively, mMIMO can provide high spectral and energy efficiency using relatively simple processing \cite{mMIMOX,lin3,lin4}. The most popular paradigm of mMIMO system is the cellular mMIMO system, where each cell has one macro base station (BS) equipped with a large number of antennas. In each cell, the BS serves a  number of user terminals (UTs) simultaneously on the same time-frequency resource. Numerous studies have validated the advantages of cellular mMIMO systems in enhancing the spectral and energy efficiency \cite{2D,youli,mMIMO3}. Nonetheless, cell-edge UTs suffer from severe performance loss in cellular mMIMO systems due to the low channel gain and the high interference from the adjacent BSs \cite{celledge}, which is an inherent problem in the cellular mMIMO system and difficult to deal with. Moreover, the handover at the cell edge may cause service interruption and delay during user mobility. With the increase of center frequency and decrease of cell radius in next generation wireless networks, these issues might become more severe \cite{precoding1}.
	
	Network mMIMO system has been proposed to enhance the quality of service of cell-edge UTs via coherent joint transmission \cite{network_MIMO,network_mMIMO,precoding1}. In the network mMIMO system, several macro BSs equipped with a large number of antennas share the data messages and channel state information (CSI) via backhaul links \cite{precoding1}. Each UT in the network is served by all the BSs and seamless services are therefore guaranteed, which not only improves the performance, but also reduces the unnecessary handover and link outage probabilities \cite{network_mMIMO}.
	Recently, the concept of network mMIMO has further evolved under the name of cell-free mMIMO, aiming to improve the quality of service for cell-edge UTs and provide a uniformly good service for all users \cite{CF1}. In cell-free mMIMO systems, the large number of transmitters, referred to as access points (APs), are geographically distributed in the network and  connected to a central processing unit (CPU) responsible for coherent transmission via backhaul links. Like network mMIMO system, each UT is served by all the APs in the cell-free mMIMO system and hence the notion of cell-edge disappears \cite{CF2}. Compared with the network mMIMO system,  the APs equipped with a much smaller number of antennas are distributed in the network much more densely in the cell-free mMIMO system. A mass of studies have borne out the superiority of the cell-free mMIMO system in boosting the spectral and energy efficiency, as well as delivering more uniform coverage to users compared to traditional cellular systems \cite{CF1,CF2,CF3}.
	However, the deployment of such a large number of APs in the real sites is a critical issue for the operators \cite{deployment}. The geographical constraints, physical obstacles, and regulatory limitations can affect the placement and density of APs, making the wide area deployment of cell-free mMIMO system a tough task \cite{CF3}, \cite{deployment}. Consequently, the network mMIMO system could be considered as a more feasible and smoother evolution of the existing system, where plenty of BSs with a large number of antennas have been deployed, and could be more practical for the next wireless generation networks for providing seamless coverage and enhancing service quality for more cell edge users over wider areas.

	The user-centric rule has been considered in many existing works \cite{UC,UC2}, and has been introduced to the cell-free mMIMO system \cite{R21,R22}. To be specific, each UT is served by a subset of the BSs that provide the best channel conditions under the user-centric rule, limiting the number of serving transmitters for each UT. Particularly, the set of BSs providing the service for the target UT is termed as the serving cluster of the UT, and equivalently, the set of UTs served by the target BS is addressed as the served group of the BS \cite{UC1,R24,R25}. In general, the dynamic serving cluster construction strategy is based on either received power or largest large-scale fading \cite{AP,R21}. In the conventional network mMIMO system, serving users with distant transmitters occupies precious power and bandwidth resources but contributes little to the performance improvement for the served UT due to the high path loss. In this regard, the integration of the network mMIMO system with the user-centric principle leads to the user-centric network (UCN) mMIMO system that we consider in this paper, which can be seen as the typical  user-centric cell-free mMIMO system when the BS density increases and the the number of antennas per BS decreases. Combining the advantages of both, the UCN mMIMO system  not only eliminates the notion of cell-edge and enhances the performances of cell-edge UTs, but also facilitates  the implementation of the network system and reduces the dimension of the precoding matrix to be designed compared with the conventional network mMIMO system \cite{UC}.

	Although the inter-cell interference can be effectively suppressed in the UCN mMIMO system, the interference in the system is still severe due to the large number of UTs in the network, making the interference management become an arduous but essential task \cite{precoding1}. Linear precoding can subdue interference and upsurge the achievable sum rate with low complexity and thus has been widely investigated \cite{precoding,sjc,yxl}.  However, the existing methods mostly involve the inverse of large dimensional matrices, increasing the computational complexity and aggravating the burden for implementation \cite{WMMSE,precoding2}. Although the introduction of the UCN can effectively reduce the computational complexity, the problem is still serious as the dimension of the matrix inversion is related to the number of transmit antennas. Even worse, higher frequency band will be explored in the future 6G wireless networks and much more antennas will be equipped at the BS side, making the problems more critical \cite{ultra_mMIMO} and necessitating the investigation of the precoder design in next generation communications.
	
	Recently, matrix manifold optimization has been widely investigated in many domains \cite{Manifold1,Manifold2,Manifold3,Manifold4} due to its ability of dealing with the equality constraints and transforming the constrained problems in Euclidean space to the unconstrained ones on manifold. Significantly, most of Riemannian methods in manifold optimization avoid the inverses of large dimensional matrices, which is of
	increasingly benefits to the next generation communications. With the combination of insights from differential geometry, optimization, and numerical analysis, matrix manifold optimization usually shows an incredible advantage in dealing with the equality constraints.   Therefore, we are motivated to investigate the precoder design for the UCN mMIMO with matrix manifold optimization.
	
	In this paper, we investigate the precoder design for UCN mMIMO downlink with matrix manifold optimization.  The main contributions of this work are the following.
We propose to combine the user-centric rule with the network mMIMO. 
On this basis, we consider the precoder design in the proposed UCN mMIMO system, whose solution space is much lower than that of the conventional network mMIMO system.
To enhance the throughput of the system, we focus on the weighted sum-rate (WSR) maximization problem and formulate a set of constraints on the problem to limit the transmit power of each BS. By proving that the precoders satisfying the constraints are on a Riemannian submanifold, we transform the constrained optimization problem in Euclidean space to an unconstrained one on the Riemannian submanifold. Then, the Riemannian ingredients, including the orthogonal projection, Riemannian gradient, retraction and vector transport, of the Riemannian submanifold are derived. With these Riemannian ingredients, Riemannian conjugate gradient (RCG) design method is proposed for solving the unconstrained optimization problem. There is no inverse of large dimensional matrix in the RCG method, which holds significant importance for the precoder design in the next generation communications.  The computational complexity of the proposed method is analyzed. 
Particularly, the acquisition of the step length involves a low computational complexity and can be ignored, demonstrating the high computational efficiency of the RCG method for precoder design in the UCN mMIMO system. Comprehensive comparisons are made between different systems, and the simulation results confirm the superiority of the UCN mMIMO system and the high efficiency of the RCG design method.
	

	The rest of this paper is organized as follows.  In \secref{sec_problem}, we first clarify the system model and formulate the precoder design problem in Euclidean space. Then the problem is reformulated on the Riemannian submanifold formed by the precoders satisfying the constraints. The Riemannian ingredients of the Riemannian submanifold needed in matrix manifold optimization are derived in \secref{sec_Riemannian_ingredients}. \secref{sec_RCG} presents the RCG design method and the complexity analysis. Simulation results  are provided in \secref{sec_Numerical_Results} to validate the superiority of the UCN mMIMO system and the  RCG design method. The conclusion is drawn in \secref{Conclusion}.
	
	\textit{Notations:} Boldface lowercase and uppercase letters represent the column vectors and matrices, respectively. We write conjugate transpose of matrix $ \bA $ as   $ \bA^{H} $ while $ \mathrm{tr}\left(\bA\right) $ and $ \det\left( \bA\right)  $ denote the matrix trace and determinant of $ \bA $, respectively. $ \Re \left\lbrace \bA \right\rbrace $ means the real part of $ \bA $ and $\mathrm{vec}\left(\bA\right)$ is the vector-version of the matrix $\bA$. Let the mathematical expectation be $ \mathbb{E}\left\{\cdot\right\}$. $ \bI_{M} $ denotes the $ M \times M $  identity matrix, whose subscript may be omitted for brevity. $ \bzero $ represents the vector or matrix whose elements are all zero. $ \diag{\ba} $ represents the diagonal matrix with $ \ba $ along its main diagonal and $ \diag{\bA} $ denotes the column vector of the main diagonal of $ \bA $. Similarly, $\bD=\mathrm{blkdiag}\left\{\bA_1,\cdots,\bA_K\right\}$ denotes the block diagonal matrix with $\bA_1,\cdots,\bA_K$ on the diagonal and $\left[\bD\right]_i$ denotes the $i$-th matrix on the diagonal, i.e., $\bA_i$. For a block matrix $\bM$, $\bM_{i,j}$ or $\left(\bM\right)_{i,j}$ denotes the $\left(i,j\right)$-th submatrix of $\bM$. $\mathrm{card}\left(\cA\right)$ denotes the  cardinality of the set $\cA$. $\cA\times \cB$ denotes the Cartesian product of the sets $\cA$ and $\cB$ and $\left(\bA,\bB\right)$ is an element in $\cA\times \cB$ with $\bA\in\cA$ and $\bB\in\cB$.  The mapping $ F $ from manifold $\cM$ to manifold $\cN$ is $F: \cM\rightarrow \cN: \bX \mapsto \bY$ denoted as $F(\bX)=\bY$.  The differential of $ F\left( \bX\right) $ is represented as $ \mathrm{D}F\left( \bX\right) $ while $ \mathrm{D}F\left( \bX\right) \left[ \bxi_{\bX} \right]  $ or $ \mathrm{D}F\left[ \bxi_{\bX} \right]  $ means the directional derivative of $ F $ at $ \bX $ along the tangent vector $ \bxi_{\bX} $.

	\section{System Model and Problem Formulation}\label{sec_problem}
	In this section, we first present the signal model of the UCN mMIMO system, where each user is served by a BS subset. Then, we formulate the WSR-maximization precoder design problem in Euclidean space with each BS having a power constraint and each UT having its own serving cluster. By proving that the precoder set satisfying the power constraints is on a Riemannian submanifold, we transform the constrained problem in Euclidean space to an unconstrained one on the Riemannian submanifold.
	\subsection{System Model}
\begin{figure}[h]
		\centering                 				
		\includegraphics[scale=0.5]{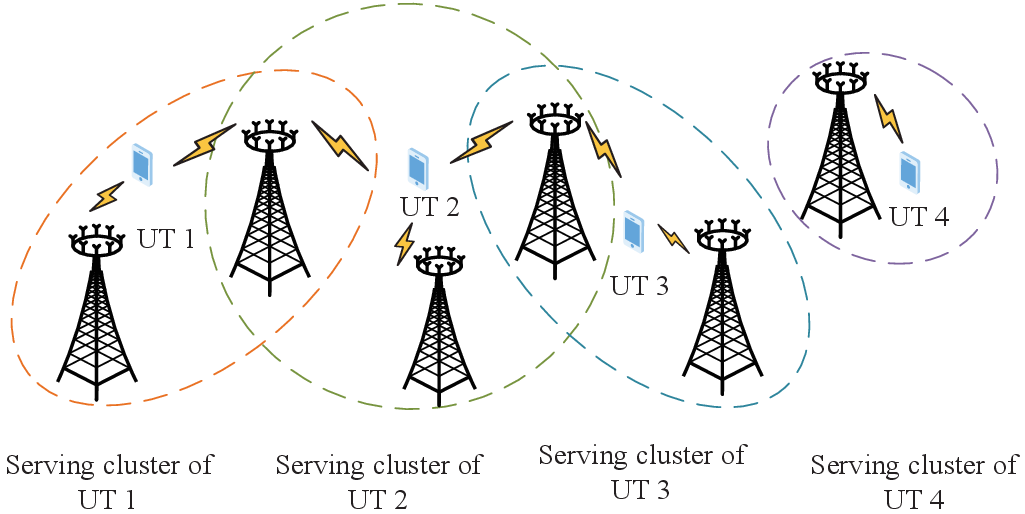}
		\caption{An illustration of the UCN mMIMO system.}
		\label{fig_configuration}
	\end{figure}

	Consider the downlink (DL) transmission in a UCN mMIMO system, where $U$ UTs are served by $B$ BSs.  The BSs are assumed to be synchronized and linked via backhaul links, which enables coherent joint transmission. Let $\cS_{B}=\left\{1,2,\cdots,B\right\}$ and $\cS_{U}=\left\{1,2,\cdots,U\right\}$ denote the sets of the BSs and the UTs, respectively. Each BS has $M_t$ transmit antennas and each UT has $M_r$ receive antennas. Each UT is served by a BS subset instead of by all the BSs, which reduces the computational burden of each BS. \figref{fig_configuration} provides an illustration of this UCN mMIMO system, where only four UTs and their serving clusters are plotted for illustrative purposes.  
	To be specific, the BSs serving UT $i$, $i\in \cS_{U}$, constitute a subset $\cB_i=\left\{ i_1, i_2, \cdots, i_{B_i} \right\}$ with $\mathrm{card}\left( \cB_i \right)=B_i$. $\cB_i$ is referred to as the serving cluster of UT $i$. The set $\cB_i, i\in\cS_{U},$ can be formed by selecting the BSs that provide the best channel conditions for UT $i$ \cite{UC}. Similarly, the UTs served by the $k$-th BS also constitute a subset $\cU_k=\left\{ k_1,k_2,\cdots,k_{U_k} \right\}$ with $\mathrm{card}\left( \cU_k \right)=U_k$, and $\cU_k$ is termed as the served group of the $k$-th BS, $k\in\cS_B$.  This user-centric rule allows each UT to have granted service without relying on the notion of cell.


	Let $\bs_{i}\in \bbC^{d_i}$ denote the $d_i$ data streams for the $i$-th UT with $\mathbb{E}\left\{\bs_{i}\bs_{i}^H\right\}=\bI_{d_i}, i\in\cS_{U},$ and $\mathbb{E}\left\{\bs_{i}\bs_{j}^H\right\}=\boldsymbol{0},  j\neq i, j\in\cS_{U}$. $\bH_{i,k}\in\bbC^{M_r\times M_t}$ is the channel matrix from the $k$-th BS to the $i$-th UT and $\bP_{i,k}\in\bbC^{M_t\times d_i}$ is the precoding matrix designed for UT $i$ by the $k$-th BS, $ k\in\cB_i$. The received signal of UT $i$ can be written as
	\begin{equation}\label{yi}
		\begin{aligned}
			\by_{i}=&\sum_{k\in\cB_i}\bH_{i,k}\bP_{i,k}\bs_i+\sum_{k\in\cB_i}\sum_{j\in \cU_k,j\neq i}\bH_{i,k}\bP_{j,k}\bs_j\\
			&+\sum_{\ell \notin \cB_i}\sum_{j\in\cU_{\ell}}\bH_{i,\ell}\bP_{j,\ell}\bs_{j}+\bz_i,
		\end{aligned}
	\end{equation}
	where $\bz_i$ is the  independent and identically distributed (i.i.d.) complex circularly symmetric Gaussian noise vector distributed as $\cC\cN\left(0,\sigma_z^2\bI_{M_r}\right)$. Denote
	\begin{equation}\label{zi}
		\begin{aligned}
			\bz_i^{\prime}&=\sum_{k\in\cB_i}\sum_{j\in \cU_k,j\neq i}\bH_{i,k}\bP_{j,k}\bs_j+\sum_{\ell \notin \cB_i}\sum_{j\in\cU_{\ell}}\bH_{i,\ell}\bP_{j,\ell}\bs_{j}+\bz_i\\
			&=\sum_{j\neq i}\sum_{\ell\in\cB_j}\bH_{i,\ell}\bP_{j,\ell}\bs_j+\bz_i
		\end{aligned}
	\end{equation}
	as the inter-user interference plus noise of UT $i$, whose covariance matrix is given by
	\begin{equation}\label{Ri}
		\begin{aligned}
			\bR_i=\sum_{j\neq i} \Bigg(\sum_{\ell\in\cB_j}\bH_{i,\ell}\bP_{j,\ell}\Bigg)\Bigg(\sum_{\ell\in\cB_j}\bH_{i,\ell}\bP_{j,\ell}\Bigg)^H+\sigma_z^2\bI_{M_r}.
		\end{aligned}
	\end{equation}
	By stacking the $\bP_{i,k}, k\in\cB_i$,  the precoder to be designed for UT $i$ can be defined as
	\begin{equation}\label{Pi}
		\begin{aligned}
			\bP_i:=\left[\bP_{i,i_{1}}^T,\cdots,\bP_{i,i_{B_i}}^T\right]^T\in\bbC^{B_iM_t\times d_i}.
		\end{aligned}
	\end{equation}	
	Note that the numbers of the rows and columns of  the precoders for different users may be different. Let  $\bH_i:=\left[\bH_{i,1},\bH_{i,2},\cdots,\bH_{i,B}\right]\in \bbC^{M_r\times BM_t}$ denote  the stacked channel matrix of UT $i$. Further,  let $\bW_i\in\bbR^{BM_t\times B_iM_t}, i\in\cS_U,$ be a block matrix composed of $B\times B_i$ submatrices, where $\left(\bW_i\right)_{i_n,n}=\bI_{M_t}$ and other submatrices are $\boldsymbol{0}_{M_t}$. Then, \eqref{yi}, \eqref{zi} and \eqref{Ri} can be rewritten as
	\begin{equation}\label{yi_new}
		\begin{aligned}
			\by_{i}&=\bH_{i}\bW_i\bP_{i}\bs_i+\bH_{i}\sum_{j\in \cS_U,j\neq i}\bW_j\bP_{j}\bs_j+\bz_i,
		\end{aligned}
	\end{equation}
	\begin{equation}\label{zi_new}
		\begin{aligned}
			\bz_i^{\prime}&=\bH_{i}\sum_{j\in \cS_U,j\neq i}\bW_j\bP_{j}\bs_j+\bz_i,
		\end{aligned}
	\end{equation}
	\begin{equation}\label{Ri_new}
		\begin{aligned}
			\bR_i&=\bH_i\sum_{j\in\cS_U,j\neq i}\bW_j\bP_j\bP_j^H\bW_j^H\bH_i^H+\sigma_z^2\bI_{M_r}.
		\end{aligned}
	\end{equation}
	
	\begin{remark}
		\rm If $B_i=B,\forall i\in\cS_U,$ the UCN mMIMO system is equivalent to the conventional network mMIMO system \cite{network_MIMO}, where the UTs are served by all the BSs. If $B_i=1,\forall i\in\cS_U,$ the system is reduced to the cellular system \cite{cellular}, where each UT is served by only one BS and the signals from other BSs are treated as inter-cell interference.
	\end{remark}
	
	\subsection{Problem Formulation in Euclidean Space}
	In this subsection, we formulate the WSR maximization precoder design problem for UCN mMIMO DL transmission in Euclidean space. For simplicity, we assume that the perfect CSI of the effective channel $\bH_{i,k}\bP_{i,k}, \forall k\in\cB_i,$ is available for the $i$-th UT via DL training \cite{YL1,zxz}. In the worst case, $\bz_i^{\prime}$ can be treated as an equivalent Gaussian noise with the covariance matrix $\bR_i$, which is assumed to be known by UT $i$. Under these assumptions, the rate of UT $i$ can be expressed as
	\begin{equation}
		\begin{aligned}
			\enspace\cR_i&\!=\!\mathrm{logdet}\left\{\bR_i\!+\!\bH_{i}\bW_i\bP_{i}\bP_{i}^H\bW_i^H\bH_{i}^H\right\}\!-\! \mathrm{logdet}\left\{\bR_i\right\}\!.
		\end{aligned}
	\end{equation}

Assuming that the transmit power of the $k$-th BS is denoted by  $P_k, k\in\cS_{B}$. Then, the power constraints of each BS can be represented by $F\left(\bP_1,\cdots,\bP_U\right)=\boldsymbol{0}$. The WSR-maximization precoder design problem can be formulated as
	\begin{equation}\label{Problem_Euclidean}
		\begin{aligned}
			\argmin{\bP_1,\cdots,\bP_U}&f\left( \bP_1,\cdots,\bP_U \right)\\
			\mathrm{s.}\mathrm{t.}\  F&\left(\bP_1,\cdots,\bP_U\right)=\boldsymbol{0},
		\end{aligned}
	\end{equation}
	where $f\left( \bP_1,\cdots,\bP_U \right)=-\sum_{i\in\cS_{U}}w_i\cR_i$ is the objective function with $w_i$ being the weighted factor of UT $i$.
	The constraint  $F\left(\bP_1,\cdots,\bP_U\right)=\boldsymbol{0}_B$  can be expressed as
	
	\begin{equation}\label{PC}
		\begin{aligned}
			&F\left(\bP_1,\cdots,\bP_U\right)=\\
			&\sum_{i\in\cU_{k}}\mathrm{tr}\left(  \bP_i^H\bW_i^H\bQ_k\bW_i\bP_i\right)-P_k=0, k\in \cS_{B},
		\end{aligned}
	\end{equation}
	where $\bQ_k=\mathrm{blkdiag}\left\{\bQ_{k,1},\cdots,\bQ_{k,B}\right\}\in\bbC^{BM_t\times BM_t}$ is a block diagonal matrix with $\bQ_{k,k}=\bI_{M_t}$ and $\bQ_{k,\ell}=\boldsymbol{0}_{M_t\times M_t}, \ell\neq k, k\in\cS_B$.

	\subsection{Problem Reformulation on Riemannian Submanifold}

	For a manifold $\mathcal{M}$, a smooth mapping $\gamma: \mathbb{R} \rightarrow \mathcal{M}: t \mapsto \gamma(t)$ is termed as a curve in $\mathcal{M}$. Let $\bX$ denote a point on $\mathcal{M}$ and $\cF_{\bX}(\mathcal{M})$ denote the set of smooth real-valued functions defined on a neighborhood of $\bX$. A \textit{tangent vector} $\bxi_{\bX}$ to manifold $\mathcal{M}$ at $\bX$ is a mapping from $\cF_{\bX}(\mathcal{M})$ to $\mathbb{R}$ such that there exists a curve $\gamma$ on $\mathcal{M}$ with $\gamma(0)=\bX$, satisfying $\bxi_{\bX} f=\left.\frac{\mathrm{d}(f(\gamma(t)))}{\mathrm{d} t}\right|_{t=0}$ for all $f \in \cF_{\bX}(\mathcal{M})$. Such a curve $\gamma$ is said to realize the tangent vector $\bxi_{\bX}$. The set of all the tangent vectors to $\cM$ at $\bX$ forms a unique and linear \textit{tangent space}, denoted by $T_{\bX}\cM$. Particularly, every vector space $\cE$ forms a \textit{linear manifold} naturally, whose  tangent space is given by $T_{\bX}\cE=\cE$.


	Besides, we can define the length of a tangent vector in $T_{\bX}\cN$ by endowing the tangent space  with an inner product $g_{\bX}^{\cN}\left(\cdot\right)$. Note that the subscript $\bX$  and the superscript $\cN$ in $g_{\bX}^{\cN}\left(\cdot\right)$ are used to distinguish the inner product of different points on different manifolds for clarity. $g_{\bX}^{\cN}\left(\cdot\right)$ is called \textit{Riemannian metric} if it varies smoothly and the manifold is called Riemannian manifold.  
	
	The product manifold is the Cartesian product of several manifolds. Let $\cM_1$ and $\cM_2$ denote two manifolds with $\bX_1\in\cM_1$ and $\bX_2\in\cM_2$, and
	\begin{equation}\label{def_product}
		\begin{aligned}
			\cM\triangleq \cM_1 \times \cM_2
		\end{aligned}
	\end{equation}
	is called the product manifold of manifold $\cM_1$ and $\cM_2$ with $\bX\triangleq\bX_1\times\bX_2\in\cM$. The tangent space of the product manifold $\cM$ is defined as
	\begin{equation}\label{tangent_product}
		\begin{aligned}
			T_{\bX}\cM=T_{\bX_1}\cM_1\times T_{\bX_2}\cM_2,
		\end{aligned}
	\end{equation}
	which is endowed with the inner product
	\begin{equation}
		\begin{aligned}
			g_{\bX}^{\cM}\left( \bxi_{\bX},\bzeta_{\bX} \right)=g_{\bX_1}^{\cM_1}\left( \bxi_{\bX_1},\bzeta_{\bX_1} \right)+g_{\bX_2}^{\cM_2}\left( \bxi_{\bX_2},\bzeta_{\bX_2} \right).
		\end{aligned}
	\end{equation}

	From the point of view of matrix manifold, the complex vector space $\bbC^{M\times N}$ forms a linear manifold naturally. So the precoding matrix $\bP_{i,k}, k\in\cB_i,$ is on the manifold $\cN_{i,k}=\bbC^{M_t\times d_i}$, whose tangent space $T_{\bP_{i,1}}\cN_{i,k}$ is still $\bbC^{M_t\times d_i}$ equipped with the Riemannian metric
	\begin{equation}
		\begin{aligned}
			g_{\bP_{i,k}}^{\cN_{i,k}}\left(\bxi_{\bP_{i,k}},\bzeta_{\bP_{i,k}}\right)=\Re\left\{\mathrm{tr}\left(\bzeta_{\bP_{i,k}}^H\bxi_{\bP_{i,k}}\right)\right\},
		\end{aligned}
	\end{equation}
	where $\bxi_{\bP_{i,k}}$ and $\bzeta_{\bP_{i,k}}$ are two tangent vectors in $T_{\bP_{i,k}}\cN_{i,k}$.
	From \eqref{def_product}, $\bP_i$ defined in \eqref{Pi} can be viewed as a point on a product manifold composed of $B_i$ manifolds defined as
	\begin{equation}\label{N_i}
		\begin{aligned}
			\cN_i:=\cN_{i,1}\times\cN_{i,2}\times \cdots\times \cN_{i,B_i}, i\in\cS_U,
		\end{aligned}
	\end{equation}
	which is equivalent to the complex vector space $\bbC^{B_iM_t\times d_i}$. From \eqref{tangent_product}, the tangent space of $\cN_i$ is given by
	\begin{equation}\label{tangent_space_Ni}
		\begin{aligned}
			T_{\bP_i}\cN_i:=T_{\bP_{i,1}}\cN_{i,1}\times T_{\bP_{i,2}}\cN_{i,2}\times \cdots \times T_{\bP_{i,B_i}}\cN_{i,B_i}, 
		\end{aligned}
	\end{equation}
	whose product Riemannian metric can be defined as
	\begin{equation}
		\begin{aligned}
			g_{\bP_i}^{\cN_i}\left(\bxi_{\bP_i},\bzeta_{\bP_i}\right)=\sum_{k\in\cB_i}\Re\left\{\mathrm{tr}\left(\bzeta_{\bP_{i,k}}^H\bxi_{\bP_{i,k}}\right)\right\}.
		\end{aligned}
	\end{equation}
	$\bxi_{\bP_i}=\left(\bxi_{\bP_{i,1}}^H,\cdots,\bxi_{\bP_{i,B}}^H \right)^H$ and $\bzeta_{\bP_i}=\left(\bzeta_{\bP_{i,1}}^H,\cdots,\bzeta_{\bP_{i,B}}^H\right)^H$ are two tangent vectors in $T_{\bP_i}\cN_i$.
	Further,  $\bP:=\left(\bP_1,\bP_2,\cdots,\bP_U\right)$ is on a product manifold defined as
	\begin{equation}
		\begin{aligned}
			\cN:=\cN_{1}\times\cN_{2}\times \cdots\times \cN_{U}.
		\end{aligned}
	\end{equation}
	The tangent space of $\cN$ is given by
	\begin{equation}\label{tangent_space_N}
		\begin{aligned}
			T_{\bP}\cN:=T_{\bP_1}\cN_1\times T_{\bP_2}\cN_2\times \cdots \times T_{\bP_U}\cN_U,
		\end{aligned}
	\end{equation}
	with the Riemannian metric
	\begin{equation}\label{Riemannian_metric}
		\begin{aligned}
			g_{\bP}^{\cN}\left(\bxi_{\bP},\bzeta_{\bP}\right)=\sum_{i\in\cS_U}g_{\bP_i}^{\cN_i}\left(\bxi_{\bP_i},\bzeta_{\bP_i}\right),
		\end{aligned}
	\end{equation}
	where $\bxi_{\bP}=\left(\bxi_{\bP_1},\cdots,\bxi_{\bP_U}\right)$
	and $\bzeta_{\bP}=\left(\bzeta_{\bP_1},\cdots,\bzeta_{\bP_U}\right)$ are two tangent vectors in $T_{\bP}\cN$. With the Riemannian metric \eqref{Riemannian_metric}, $\cN$ is a Riemannian product manifold.  Let
	\begin{equation}\label{M_definition}
		\begin{aligned}
			\cM:=\left\{ \bP\in\cN\mid  F\left(\bP\right)=\boldsymbol{0}_B \right\}
		\end{aligned}
	\end{equation}
	denote the set of the precoders  satisfying \eqref{PC}. Then, we have the following theorem.
	\begin{theorem}\label{theorem_submanifold}
		$\cM$ defined in \eqref{M_definition} forms a Riemannian  submanifold of $\cN$ with the Riemannian metric
		\begin{equation}\label{Riemannian_metric_M}
			\begin{aligned}
				g_{\bP}^{\cM}\left(\bxi_{\bP},\bzeta_{\bP}\right) = g_{\bP}^{\cN}\left(\bxi_{\bP},\bzeta_{\bP}\right),  \bxi_{\bP},\bzeta_{\bP}\in T_{\bP}\cM.
			\end{aligned}
		\end{equation}
		\begin{proof}
			See the proof in \appref{app_submanifold}.
		\end{proof}
	\end{theorem}
	
			%
	It is worth emphasizing that that $\bP\in\cM$ satisfying $F\left(\bP\right)=\boldsymbol{0}_B$ is still on $\cN$.  So $\bxi_{\bP}$ and $\bzeta_{\bP}$ on the right hand side of \eqref{Riemannian_metric_M}  are viewed as elements in $T_{\bP}\cN$. From \thref{theorem_submanifold}, we can reformulate the constrained problem \eqref{Problem_Euclidean} as an unconstrained one on $\cM$ as 
	\begin{equation}\label{Problem_manifold}
		\begin{aligned}
			\argmin{\bP\in\cM}f\left(\bP\right).
		\end{aligned}
	\end{equation}

	\section{Riemannian Ingredients for Precoder Design }\label{sec_Riemannian_ingredients}
	In this section, we derive all the Riemannian ingredients needed for solving \eqref{Problem_manifold} in manifold optimization, including orthogonal projection, Riemannian gradient, retraction and vector transport.

	\subsection{Orthogonal Projection}
\begin{figure}[t]
		\centering                 				
		\includegraphics[scale=0.6]{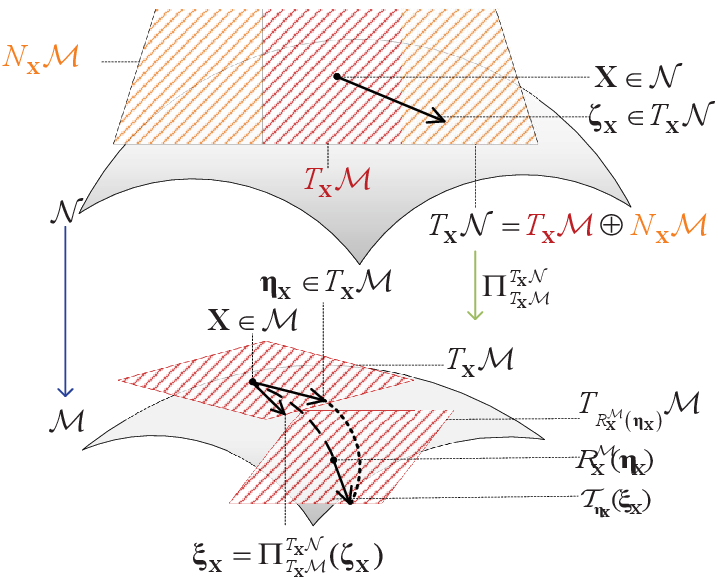}
		\caption{ Geometric interpretation of orthogonal projection, retraction and vector transport.}
		\label{fig_manifold_tangent}
	\end{figure}

	With the Riemannian metric $g_{\bP}^{\cN}\left(\cdot\right)$, the tangent space $T_{\bP}\cN$ at $\bP\in\cN$ defined in \eqref{tangent_space_N} can be decomposed into two orthogonal subspaces as
	\begin{equation}\label{decomposition_tangent_space}
		T_{\bP}\cN=T_{\bP}\cM \oplus N_{\bP}\cM,
	\end{equation}
	where $T_{\bP}\cM$ and  $N_{\bP}\cM$ are the tangent space and the normal space of $\cM$  at $\bP\in\cM$, respectively. For geometric understanding, \figref{fig_manifold_tangent} is a simple illustration.
	
	From \eqref{PC} and \eqref{M_definition}, it is easy to verify that  $\cM$ is defined as a level set of a constant-rank function $F$ \cite{Absil2009}. In this case, $	T_{\bP}{\cM}$ is the kernel of the differential of $F$ and a subspace of $	T_{\bP}{\cN}$ defined as \cite{Absil2009}
	\begin{equation}\label{tangent_space}
		\begin{aligned}
			T_{\bP}\cM=\mathrm{ker}\left( \mathrm{D}F\left(\bP\right) \right).
		\end{aligned}
	\end{equation}
	Recall that $F:\cN \rightarrow \bbR^B$ defined in \eqref{PC} is a smooth mapping from manifold $\cN$ to manifold $\overline{\cM}$, where $\cN$ and $\overline{\cM}$ are the manifolds formed by the vector space $\bbC^{BM_t\times \sum_{i=1}^{U}d_i}$ and $\bbR^B$, respectively.  $\mathrm{D}F\left(\bP\right)\left[\cdot\right]$  is thus a linear mapping from $T_{\bP}\cN$ to $T_{\bY}\overline{\cM}$, where $\bY=F\left(\bP\right)=\boldsymbol{0}_B$ is a point on $\overline{\cM}$. Hence, $\bxi_{\bY}=\mathrm{D}F\left(\bP\right)\left[\bxi_{\bP}\right]$ is a tangent vector to $\overline{\cM}$ at $\bY$, i.e.,  an element in $T_{\bY}\overline{\cM}$. 
	In particular, as $\cN=\bbC^{BM_t\times \sum_{i\in\cS_{U}}d_i}$ and $\overline{\cM}=\bbR^B$ are  both linear manifolds, $\mathrm{D}F\left(\bP\right)$ will be reduced to the classical directional derivative
	\begin{equation}\label{DFX}
		\begin{aligned}
			\mathrm{D}F\left(\bP\right)\left[ \boldsymbol{\bxi}_{\bP} \right] = \lim_{t\rightarrow 0}\frac{F\left(\bP+t\boldsymbol{\bxi}_{\bP}\right)-F\left(\bP\right)}{t}.
		\end{aligned}
	\end{equation}
	From \eqref{tangent_space} and \eqref{DFX}, $T_{\bP}\cM$ is given by
	
	\begin{equation}\label{tangent_space_M}
		\begin{aligned}
			T_{\bP}\cM= \Big\{ \bxi_{\bP}\in T_{\bP}\cN \mid \sum_{i\in\cU_{k}}\mathrm{tr}\Big(  \bxi_{\bP_i}^H\bW_i^H\bQ_k\bW_i\bP_i\\
			+\bP_i^H\bW_i^H\bQ_k\bW_i\bxi_{\bP_i}\Big)=0, k\in\cS_{B}\Big\}.
		\end{aligned}
	\end{equation}
	To obtain the elements in $T_{\bP}{\cM}$, we can obtain the corresponding elements in $T_{\bP}{\cN}$ first and turn to the \textit{orthogonal projection}.  The normal space $N_{\bP}\cM$ is the orthogonal complement of $T_{\bP}\cM$ and thus can be expressed as
	\begin{equation}\label{normal_space_M}
		\begin{aligned}
			N_{\bP}\cM=&\Big\{ \bzeta_{\bP}\in T_{\bP}\cN \mid g_{\bP}^{\cN}\left(\bxi_{\bP},\bzeta_{\bP}\right)=0,\forall \bxi_{\bP}\in T_{\bP}\cM \Big\}\\
			=&\Big\{\Big( \sum_{\ell\in\cB_1}\mu_\ell\bW_1^H \bQ_\ell\bW_1\bP_1,\cdots,\\
			&\sum_{\ell\in\cB_U}\mu_\ell\bW_U^H \bQ_\ell\bW_U\bP_U \Big)\mid \mu_\ell\in\bbR\Big\}.
		\end{aligned}
	\end{equation}
	With \eqref{decomposition_tangent_space}, any $\bxi_{\bP}\in T_{\bP}\cN$ can be decomposed into two orthogonal tangent vectors as
	\begin{equation}\label{decomposition_xi}
		\bxi_{\bP}=\bPi^{T_{\bP}\cN}_{T_{\bP}\cM}\left(\bxi_{\bP}\right)+\bPi^{T_{\bP}\cN}_{N_{\bP}\cM}\left(\bxi_{\bP}\right),
	\end{equation}
	where $\bPi^{T_{\bP}\cN}_{T_{\bP}\cM}\left(\bxi_{\bP}\right)$ and $\bPi^{T_{\bP}\cN}_{N_{\bP}\cM}\left(\bxi_{\bP}\right)$ represent the orthogonal projections of $\bxi_{\bP}$ onto $T_{\bP}\cM$ and $N_{\bP}\cM$, respectively.
	\begin{lemma}\label{lemma_projection}
		For any $\bxi_{\bP}\in T_{\bP}\cN$, the orthogonal projection $\bPi^{T_{\bP}\cN}_{T_{\bP}\cM}\left(\bxi_{\bP}\right)$ is given by
		\begin{equation}\label{orthogonal_projection}
			\begin{aligned}
				\bPi^{T_{\bP}\cN}_{T_{\bP}\cM}\left(\bxi_{\bP}\right)=\Big(& \bxi_{\bP_1}-\sum_{\ell\in\cB_1}\mu_\ell \bW_1^H\bQ_\ell\bW_1\bP_1,\cdots,\\
				&\bxi_{\bP_U}-\sum_{\ell\in\cB_U}\mu_\ell \bW_U^H\bQ_\ell\bW_U\bP_U \Big),
			\end{aligned}
		\end{equation}
		where
		\begin{equation}
			\mu_\ell=\frac{1}{P_{\ell}}\sum_{i\in\cU_{\ell}}\Re\left\{ \mathrm{tr}\left( \bP_i^H\bW_i^H\bQ_\ell\bW_i
			\bxi_{\bP_i} \right) \right\}.
		\end{equation}
		\begin{proof}
			See the proof in \appref{app_projection}.
		\end{proof}
	\end{lemma}

	\subsection{Riemannian Gradient}
	
	The set of all tangent vectors to $\cM$ is called \textit{tangent bundle} denoted by $T\cM$, which itself is a smooth manifold. A \textit{vector field} $\bxi$ on manifold $\cM$ is a smooth mapping from $\cM$ to $T\cM$ that assigns to each point $\bP\in\cM$ a tangent vector $\bxi_{\bP}\in T_{\bP}\cM$. Denote $\mathrm{grad}_{\cM}f$ as the vector field of the \textit{Riemannian gradient}. For the smooth real-valued function $f$ on  Riemannian submanifold $\cM$, the Riemannian gradient of $f$ at $\bP$, denoted by $\mathrm{grad}_{\cM}f(\bP)$, is defined as the unique element in $T_{\bP}{\cM}$ that satisfies
	\begin{equation}\label{RawGradient}
		\begin{aligned}
			g_{\bP}^{\cM}\left( \mathrm{grad}_{\cM}f(\bP),\boldsymbol{\bxi}_{\bP} \right) = \mathrm{D}f\left(\bP\right)\left[ \boldsymbol{\bxi}_{\bP} \right], \forall \bxi_{\bP}\in T_{\bP}\cM.
		\end{aligned}
	\end{equation}
	Note that, since $\mathrm{grad}_{\cM}f(\bP)\in T_{\bP}{\cM}$, we can derive the Riemannian gradient $\mathrm{grad}_{\cM}f\left(\bP\right)$ in $T_{\bP}\cM$ by projecting the Riemannian gradient $\mathrm{grad}_{\cN}f\left(\bP\right)$ in $T_{\bP}\cN$ onto $T_{\bP}\cM$, which will play a significant role in obtaining the search direction in optimization. Denoting
	\begin{equation}\label{C_i} 
		\bC_i=\left(\bI_{d_i}+\bP_i^H\bW_i^H\bH_i^H\bR_i^{-1}\bH_i\bW_i\bP_i\right)^{-1},
	\end{equation}
	we have the following theorem.
	\begin{theorem}\label{theo_gradient}
		The Euclidean gradient of $f\left(\bP\right)$  is given by
		\begin{equation}\label{gradient_N}
			\begin{aligned}
				\mathrm{grad}_{\cN}f\left(\bP\right)=\left( \mathrm{grad}_{\cN_1}f\left(\bP_1\right),\cdots,\mathrm{grad}_{\cN_U}f\left(\bP_U\right) \right),
			\end{aligned}
		\end{equation}
		where
		\begin{equation}
			\begin{aligned}
				\mathrm{grad}_{\cN_i}&f\left(\bP_i\right)=\\
				&\left( \mathrm{grad}_{\cN_{i,1}}f\left(\bP_{i,1}\right)^T,\cdots,\mathrm{grad}_{\cN_{i,B}}f\left(\bP_{i,B}\right)^T \right)^T.
			\end{aligned}	
		\end{equation}
		To be specific, 
		\begin{equation}\label{gradient_Ni}
			\begin{aligned}
				&\mathrm{grad}_{\cN_{i,k}}f\left(\bP_{i,k}\right)=-2\big( w_i\bH_{i,k}^H\bR_i^{-1}\bH_{i}\bW_i\bP_{i}\bC_i \\
				&-\sum_{j \neq i}^{U}w_j\bH_{j,k}^H\bR_{j}^{-1}\bH_{j}\bW_j\bP_{j}\bC_{j}\bP_{j}^H\bW_j^H\bH_{j}^H\bR_{j}^{-1}\bH_{j}\bW_i\bP_{i}\big)
			\end{aligned}
		\end{equation}
		in $ T_{\bP_{i,k}}\cN_{i,k}=\bbC^{M_t\times d_i}$ is the Euclidean gradient of UT $i$ served by the $k$-th BS for $k\in\cB_i, i\in\cS_{U}$. The Riemannian gradient of $f\left(\bP\right)$ in $ T_{\bP}\cM$  is given by
		\begin{equation}\label{gradient_M}
			\begin{aligned}
				\mathrm{grad}_{\cM}&f\left(\bP\right)=\\
				&\Big(\mathrm{grad}_{\cN_1}f\left(\bP_1\right)-\sum_{k\in\cB_1}\lambda_k\bW_1^H\bQ_k\bW_1\bP_1,\cdots,\\
				&\mathrm{grad}_{\cN_U}f\left(\bP_U\right)-\sum_{k\in\cB_U}\lambda_k\bW_U^H\bQ_k\bW_U\bP_U\Big),
			\end{aligned}
		\end{equation}
		where
		\begin{equation}
			\begin{aligned}
				\lambda_k=\frac{1}{P_{k}}\sum_{i\in\cU_k}\Re\Big\{ \mathrm{tr}\left( \bP_i^H\bW_i^H\bQ_k\bW_i
				\mathrm{grad}_{\cN_i}f\left(\bP_i\right) \right) \Big\}.
			\end{aligned}
		\end{equation}
		\begin{proof}
			See the proof in \appref{app_gradient}.
		\end{proof}
	\end{theorem}
	It is worth noting that only $\mathrm{grad}_{\cN_{i,k}}f\left(\bP_{i,k}\right),  k\in\cB_{i},$ are computed,  which will decrease the computational complexity of $\mathrm{grad}_{\cM}f\left(\bP\right)$ compared with the conventional network systems.

	\subsection{Retraction}
	For a nonlinear manifold, the notion of moving along the tangent vector while remaining on the manifold and preserving the search direction is generalized by \textit{retraction}. The retraction $R_{\bP}^{\cM}\left(\cdot\right)$ is a smooth mapping from $T_{\bP} \cM$ to $\cM$ \cite[Definition 4.1.1]{Absil2009} and builds a bridge between the linear $T_{\bP} \cM$ and the nonlinear $\cM$.  For the Riemannian submanifold $\cM$, a computationally efficient retraction $R_{\bP}^{\cM}\left(\bxi_{\bP}\right), \bxi_{\bP}\in T_{\bP} \cM,$ can be computed by projecting $\left(\bP+\bxi_{\bP}\right)\in\cN$ back to the manifold $\cM$ \cite[Section 4.1.1]{Absil2009}.
	
	\begin{theorem}\label{theo_retraction}
		Let
		\begin{equation}
			\begin{aligned}
				\gamma_k&=\frac{\sqrt{P_k}}{\sqrt{\sum_{i\in\cU_k}\mathrm{tr}\left(\left(\bP_i+\bxi_{\bP_i}\right)^H\bW_i^H\bQ_k\bW_i\left(\bP_i+\bxi_{\bP_i}\right)\right)}},\\
				\bGamma_i&=\mathrm{blkdiag}\left(\gamma_{i_1}\bI_{M_t},\gamma_{i_2}\bI_{M_t},\cdots,\gamma_{i_{B_i}}\bI_{M_t}\right),  i\in\cS_U.
			\end{aligned}
		\end{equation}
		Then, the retraction from $T_{\bP}\cM$ to $\cM$ is given by
		\begin{equation}\label{retraction}
			\begin{aligned}
				R_{\bP}^{\cM}\left(\bxi_{\bP}\right): &\ T_{\bP}\cM\rightarrow\cM\\
				: &\ \bxi_{\bP}\mapsto \left(\bGamma_1\left(\bP_1+\bxi_{\bP_1}\right),\cdots,\bGamma_U\left(\bP_U+\bxi_{\bP_U}\right)\right),
			\end{aligned}
		\end{equation}
		where $\bxi_{\bP}$ is usually a search direction.
		\begin{proof}
			The result can be easily verified by substituting $R_{\bP}^{\cM}\left(\bxi_{\bP}\right)$ to \eqref{PC} and the proof is omitted.
		\end{proof}
	\end{theorem}
	
	\begin{remark}
		\rm	Let $\bar{\bP}_k:=\left(\bP_{k_1,k},\bP_{k_2,k},\cdots,\bP_{k_{U_k},k}\right)\in\bbC^{M_t\times\sum_{i\in\cU_{k}}d_i}$ be the stacked precoder matrix of all the users  served by the $k$-th BS. 
		From the perspective of geometry, \eqref{retraction} normalizes the transmit power of each BS and forces $\mathrm{vec}\left(\bar{\bP}_k\right)$ to stay on a sphere of radius $\sqrt{P_k},  k\in\cS_B$.
	\end{remark}

	\subsection{Vector Transport}
	It is obvious that the Riemannian submanifold $\cM$ is nonlinear as $(\bP_1+\bP_2)\notin \cM$, where $\bP_1,\bP_2\in\cM$.  The addition of tangent vectors in different tangent spaces is not straightforwardly in $\cM$ as the tangent spaces at different points on $\cM$ are different. \textit{Vector transport} denoted by $\cT_{\boldsymbol{\eta}_{\bP}}^{\cM}\left(\bxi_{\bP}\right)\in T_{R_{\bP}^{\cM}\left(\boldsymbol{\eta}_{\bP}\right)}\cM$  is thus introduced  to transport a tangent vector $\bxi_{\bP}$ from a point $\bP\in\cM$ to another point $R_{\bP}^{\cM}\left(\boldsymbol{\eta}_{\bP}\right)\in\cM$. For $\cM$, the vector transport can be obtained according to the following theorem.
	
	\begin{theorem}\label{theo_vector_transport}
		Let $\bP^{\mathrm{new}}=R_{\bP}^{\cM}\left(\boldsymbol{\eta}_{\bP}\right)$. Then, the vector transport on $\cM$ is given by
		\begin{equation}\label{vector_transport}
			\begin{aligned}
				\cT_{\boldsymbol{\eta}_{\bP}}^{\cM}\left(\bxi_{\bP}\right)=&\bPi^{T_{\bP}\cM}_{T_{\bP^{\mathrm{new}}}\cM}\left(\bxi_{\bP}\right)\\
				=&\Big(\bxi_{\bP_1}-\sum_{\ell\in\cB_1}\rho_{\ell}\bW_1^H\bQ_{\ell}\bW_1\bP^{\mathrm{new}}_1,\cdots,\\
				&\bxi_{\bP_U}-\sum_{\ell\in\cB_U}\rho_{\ell}\bW_U^H\bQ_{\ell}\bW_U\bP^{\mathrm{new}}_U\Big),
			\end{aligned}
		\end{equation}
		where
		\begin{equation}
			\rho_{\ell}=\frac{1}{P_{\ell}}\sum_{i\in\cU_{\ell}}\Re\left\{\mathrm{tr}\left(\left(\bP^{\mathrm{new}}_i\right)^H\bW_i^H\bQ_\ell\bW_i\bxi_{\bP_i}\right)\right\}, \ell \in \cS_B.
		\end{equation}
		\begin{proof}
			See the proof in \appref{app_vector_transport}.
		\end{proof}
	\end{theorem}

	\section{Riemannian Conjugate Gradient Precoder Design}\label{sec_RCG}
	In this section, we first revisit the conventional conjugate gradient method in Euclidean space and then introduce the RCG method for precoder design in the UCN mMIMO with the Riemannian ingredients derived in \secref{sec_Riemannian_ingredients}. The proposed design obviates the need for inverses of large dimensional matrices, which is beneficial for practice. The computational complexity of the proposed method is analyzed, showing the computational efficiency of our precoder design.
	\subsection{Conventional Conjugate Gradient Method}
	Line search is one of the most well-known strategies for unconstrained optimization of smooth functions in Euclidean space \cite{numerical}. In the line search strategy, the algorithm chooses a direction and searches along this direction from the current point to a new point with a lower objective function value.  For notational clarity, any variable with the superscript $n$ represents the variable in the $n$-th iteration of the line search method. The conventional update formula  is given by
	\begin{equation}\label{update_Euclidean_out}
		\begin{aligned}
			\bP^{n+1}=\bP^{n}+\alpha^n\boldsymbol{\eta}^n,
		\end{aligned}
	\end{equation}
	where $\alpha\in\bbR$ and $\boldsymbol{\eta}$ are the step length and search direction, respectively. If $\boldsymbol{\eta}$ is chosen as the negative gradient of the objective function during the iteration, \eqref{update_Euclidean_out} is the update formula of the steepest gradient descent method, which is efficient but converges slowly. Conjugate gradient method accelerates the convergence rate by modifying the search direction, which is given by
	\begin{equation}\label{direction_Euclidean}
		\boldsymbol{\eta}^{n}=-\mathrm{grad}_{\cN}f\left(\bP^n\right)+\beta^{n}\boldsymbol{\eta}^{n-1},
	\end{equation}  
	where $\beta^n\in\bbR$ is a scalar.
	During each iteration, a limited number of trial step lengths are generated to search for an  effective point along the search direction $\boldsymbol{\eta}^n$ that decreases the value of the objective function \cite{numerical}. Let the superscript pair $\left(n,m\right)$ represent the $m$-th inner iteration of searching for the step length during the $n$-th outer iteration. The conventional update formula for searching for the step length is given by 
	\begin{equation}\label{update_Euclidean_in}
		\begin{aligned}
			\bP^{n,m+1}=\bP^{n}+\alpha^{n,m}\boldsymbol{\eta}^n.
		\end{aligned}
	\end{equation}
	\eqref{update_Euclidean_in} is repeated until an efficient $\alpha^{n}=\alpha^{n,m}$ is obtained that ensures an enough decrease of the objective function with $\bP^{n+1}=\bP^{n,m+1}$. \eqref{update_Euclidean_out} is repeated  until a good enough $\bP=\bP^{n+1}$ is reached. 
	\subsection{Riemannian Conjugate Gradient Precoder Design}
	For the optimization on the manifold, the conventional update formula in \eqref{update_Euclidean_out} is not suitable for nonlinear manifold as $\left(\bP^{n}+\alpha^n\boldsymbol{\eta}^n\right)$ is not necessary on the manifold. Retraction derived in \thref{theo_retraction} is utilized to keep $\bP^{n+1}$ on the manifold and preserve the search direction. From \thref{theo_retraction}, the update formula on $\cM$ is given by
	\begin{equation}\label{update_manifold_out}
		\begin{aligned}
			\bP^{n+1}=\Big(\bGamma^n_1\left(\bP_1^{n}+\alpha^n\boldsymbol{\eta}_1^n\right),\cdots,\bGamma^n_U\left(\bP_U^{n}+\alpha^n\boldsymbol{\eta}_U^n\right)\Big)
		\end{aligned}
	\end{equation}
	with $\boldsymbol{\eta}=\left(\boldsymbol{\eta}_1,\boldsymbol{\eta}_2,\cdots,\boldsymbol{\eta}_U\right)\in T_{\bP}\cM$ being the search direction.
	To be specific,  \eqref{update_manifold_out} can be rewritten as 
	\begin{equation}\label{update_user}
		\begin{aligned}
			\bP^{n+1}_{i}&=\bGamma^n_i\left(\bP^n_{i}+\alpha^n\boldsymbol{\eta}_{i}^n\right), i\in \cS_U.
		\end{aligned}
	\end{equation}
	More specifically, $\boldsymbol{\eta}^n_i$ can be written as  $\Big(\left(\boldsymbol{\eta}^n_{i,1}\right)^T,\left(\boldsymbol{\eta}^n_{i,2}\right)^T,$ $\cdots,\left(\boldsymbol{\eta}^n_{i,B}\right)^T\Big)^T$from \eqref{tangent_space_Ni} and  \eqref{update_user} can be further refined to
	\begin{equation}\label{update_single}
		\begin{aligned}
			\bP^{n+1}_{i,k}&=\gamma_{k}^n\left(\bP^n_{i,k}+\alpha^n\boldsymbol{\eta}_{i,k}^n\right), i\in \cS_U, k\in \cB_i.
		\end{aligned}
	\end{equation}
	With the assistance of the vector transport derived in \thref{theo_vector_transport}, the search direction \eqref{direction_Euclidean} can be adjusted as
	\begin{equation}\label{eta}
		\boldsymbol{\eta}^{n+1}=-\mathrm{grad}_{\cM}f\left(\bP^n\right)+\beta^n	\cT_{\alpha^n\boldsymbol{\eta}^n}\left(\boldsymbol{\eta}^n\right),
	\end{equation}
	where $\beta^n\in\bbR$ is the RCG update parameter with several alternatives  that yield different nonlinear RCG methods \cite{sato}.  $\beta^n$ is chosen as the modified Polak and Ribi\`{e}re parameter (PRP) to avoid jamming and is given by
	\begin{equation}
		\begin{aligned}
			\beta^n=\max\left(0,\min\left( \beta^n_{\mathrm{PRP}},\beta^n_{\mathrm{FR}} \right)\right),
		\end{aligned}
	\end{equation} 
	where
	\begin{equation}
		\begin{aligned}
			\setlength\abovedisplayskip{3pt}
			\setlength\belowdisplayskip{3pt}
			\beta^n_{\mathrm{FR}}=\frac{g^{\cM}_{\bP^n}\big(\mathrm{grad}_{\cM}f\left(\bP^n\right),\mathrm{grad}_{\cM}f\left(\bP^n\right) \big)} {g^{\cM}_{\bP^{n-1}}\big(\mathrm{grad}_{\cM}f\left(\bP^{n-1}\right),\mathrm{grad}_{\cM}f\left(\bP^{n-1}\right)  \big)}.
		\end{aligned}
	\end{equation}
	Let $\boldsymbol{\nu}^n=\mathrm{grad}_{\cM}f\left(\bP^n\right)-\cT_{\alpha^{n-1}\boldsymbol{\eta}^{n-1}}^{\cM}\big(\mathrm{grad}_{\cM}f\left(\bP^{n-1}\right)\big)$, $\beta^n_{\mathrm{PRP}}$ is given by
	\begin{equation}
		\setlength\abovedisplayskip{3pt}
		\setlength\belowdisplayskip{3pt}
		\begin{aligned}
			\beta^n_{\mathrm{PRP}}=\frac{g^{\cM}_{\bP^n}\big(\mathrm{grad}_{\cM}f\left(\bP^n\right),\boldsymbol{\nu}^n \big)} {g^{\cM}_{\bP^{n-1}}\big(\mathrm{grad}_{\cM}f\left(\bP^{n-1}\right),\mathrm{grad}_{\cM}f\left(\bP^{n-1}\right)  \big)}.
		\end{aligned}
	\end{equation}
	Define $\bV_{i,j,\ell}:=\bH_{i,\ell}\bP_{j,\ell}\in\bbC^{M_r\times d_i}$ and $\bU_{i,j,\ell}:=\bH_{i,\ell}\boldsymbol{\eta}_{j,\ell}\in\bbC^{M_r\times d_i}, i,j\in\cS_{U}, \ell\in\cB_j$. $\bV_{i,j,\ell}$ in the $(n+1)$-th iteration can be written as
	\begin{equation}\label{V_single}
		\begin{aligned}
			\bV_{i,j,\ell}^{n+1}&=\gamma_{\ell}^{n}\bH_{i,\ell}\left(\bP^{n}_{j,\ell}+\alpha^{n}\boldsymbol{\eta}_{j,\ell}^n\right)\\
			&=\gamma_{\ell}^{n}\left(\bV_{i,j,\ell}^n+\alpha^n\bU_{i,j,\ell}^n\right).
		\end{aligned}
	\end{equation}
	Further, define $\bV_{i,j}:=\bH_i\bW_j\bP_j\in\bbC^{M_r\times d_j}$ and $\bU_{i,j}:=\bH_i\bW_j\boldsymbol{\eta}_j\in\bbC^{M_r\times d_j}$. $\bV_{i,j}$ and $\bU_{i,j}$ in the $(n+1)$-th iteration can be expressed as
	\begin{subequations}
		\begin{align}
			\bV_{i,j}^{n+1}&=\sum_{\ell\in\cB_j}\bV_{i,j,\ell}^{n+1}=\sum_{\ell\in\cB_j}\gamma_{\ell}^{n}\left(\bV_{i,j,\ell}^n+\alpha^n\bU_{i,j,\ell}^n\right)\label{V_user},\\
			\bU_{i,j}^{n+1}&=\sum_{\ell\in\cB_j}\bU_{i,j,\ell}^{n+1}\label{U_user},
		\end{align}
	\end{subequations}
	respectively.
	With $\bV_{i,j}$, the covariance matrix in the $n$-th iteration can be rewritten as
	\begin{equation}\label{R_i}
		\begin{aligned}
			\bR_i^n=\sum_{j\neq i,j\in\cS_U}\bV_{i,j}^n\left(\bV_{i,j}^n\right)^H+\sigma_z^2\bI_{M_r}\in\bbC^{M_r\times M_r}.
		\end{aligned}
	\end{equation}
	Then the Euclidean gradient of UT $i$ served by the $k$-th BS,  $k\in\cB_i,$ in the $n$-th iteration can be rewritten as
	\begin{equation}\label{gradient_Ni_n}
		\begin{aligned}
			&\mathrm{grad}_{\cN_{i,k}}f\left(\bP_{i,k}^n\right)=\\
			& -2w_i\bH_{i,k}^H\left(\bR_i^n\right)^{-1}\bV_{i,i}^n\left(\bI_{d_i}+\left(\bV_{i,i}^n\right)^H\left(\bR_i^n\right)^{-1}\bV_{i,i}^n\right)^{-1}+ \\
			&2\sum_{ j \neq i}^{U}w_j\bH_{j,k}^H\left(\bR_j^n\right)^{-1}\bV_{j,j}^n\left(\bI_{d_j}+\left(\bV_{j,j}^n\right)^H\left(\bR_j^n\right)^{-1}\bV_{j,j}^n\right)^{-1}\\
			&\times\left(\bV_{j,j}^n\right)^H\left(\bR_j^n\right)^{-1}\bV_{j,i}^n,
		\end{aligned}
	\end{equation}
	which is only related to $\bH_{i,k}$ and $\bV_{i,j,k}^n,  i,j\in\cS_U, k\in\cB_i$. Like \eqref{update_single}, the update formula of searching for the step length in manifold optimization is adjusted as
	\begin{equation}\label{retraction_in}
		\begin{aligned}
			\bP^{n,m+1}_{i,k}&=\gamma_{k}^{n,m+1}\left(\bP^n_{i,k}+\alpha^{n,m+1}\boldsymbol{\eta}_{i,k}^n\right), k\in \cB_i.
		\end{aligned}
	\end{equation}
	For efficiency of computation, the step length can be obtained by the backtracking method \cite{Absil2009}. During the iteration for searching the step length in the $n$-th outer iteration, $\bP^n$ and $\boldsymbol{\eta}^n$ are fixed and the objective function can be viewed as a function of $\alpha$ and is given by
	\begin{equation}
		\phi\left(\alpha\right)=f\left(R^{\cM}_{\bP^n}\left(\alpha\boldsymbol{\eta}^{n}\right)\right).
	\end{equation}
	To be specific, the objective function in the $\left(n,m+1\right)$-th iteration is determined by $\alpha^{n,m}$ and can be written as
	\begin{equation}\label{phi_nm}
		\phi\left(\alpha^{n,m}\right)=\sum_{i=1}^{U}w_i\cR_i^{n,m+1},
	\end{equation}
	where
	\begin{equation}\label{R_nm}
		\begin{aligned}
			&\cR_i^{n,m+1}=\mathrm{logdet}\Bigg(\sum_{j\in\cS_U}\bV_{i,j}^{n,m+1}\left(\bV_{i,j}^{n,m+1}\right)^H+\sigma_z^2\bI_{M_r}\Bigg)-\\
			&\mathrm{logdet}\Bigg(\sum_{j\neq i,j\in\cS_U}\bV_{i,j}^{n,m+1}\left(\bV_{i,j}^{n,m+1}\right)^H+\sigma_z^2\bI_{M_r}\Bigg)
		\end{aligned}
	\end{equation}
	with
	\begin{equation}\label{R_i_nm}
		\begin{aligned}
			\bR_i^{n,m+1}=\sum_{j\neq i,j\in\cS_U}\bV_{i,j}^{n,m+1}\left(\bV_{i,j}^{n,m+1}\right)^H+\sigma_z^2\bI_{M_r}.
		\end{aligned}
	\end{equation}
	\eqref{R_nm} is determined by the low dimensional matrix  $\bV_{i,j}^{n,m+1}\in\bbC^{M_r\times d_i}$, which can be directly obtained from
	\begin{equation}\label{V_in}
		\begin{aligned}
			\bV_{i,j}^{n,m+1}&=\sum_{\ell\in\cB_j}\bV_{i,j,\ell}^{n,m+1}\\
			&=\sum_{\ell\in\cB_j}\gamma_{\ell}^{n,m}\left(\bV_{i,j,\ell}^{n}+\alpha^{n,m}\bU_{i,j,\ell}^{n}\right).
		\end{aligned}
	\end{equation}
	The RCG method for precoder design in the UCN mMIMO system is provided in \alref{RCG}, where $r$ and $c$ are typically chosen as $0.5$ and $10^{-4}$, respectively. The RCG method can converge to a stationary point and the convergence behavior analyses are shown in \cite{sato,Boumal2020}.
	\begin{algorithm}[h]  
		\caption{RCG method for precoder design in the UCN mMIMO system} 
		\label{RCG}  
		\renewcommand{\algorithmicensure}{\textbf{Output:}}
		\begin{algorithmic}[1] 
			\REQUIRE Riemannian submanifold $\cM$; Riemannian metric $g_{\bP}^{\cM}\left(\cdot\right)$; Real-valued function $f$;  \\
			Retraction $R_{\bP}^{\cM}\left(\cdot\right)$; Vector transport $\cT_{\boldsymbol{\eta}}^{\cM}\left(\cdot\right)$; initial step length $\alpha^0>0$; $r\in \left(0,1\right)$; $c\in \left(0,1\right)$
			\renewcommand{\algorithmicrequire}{\textbf{Input:}}
			\REQUIRE Initial point $\bP^0$;
			\REPEAT
			\STATE Get $\bV_{i,j,\ell}^n$ with \eqref{V_single} and $\bV_{i,j}^n$ with \eqref{V_user} for $ i,j\in\cS_{U},\ell\in\cB_j$.
			\STATE  Compute Euclidean gradient $\mathrm{grad}_{\cN}f(\bP^n)$ with \eqref{gradient_Ni_n}. 
			\STATE Get Riemannian gradient $\mathrm{grad}_{\cM}f(\bP^{n})$ with \eqref{gradient_M}.
			\STATE Update the search direction $\boldsymbol{\eta}^{n}$ with \eqref{eta} and compute $\bU_{i,j}^n, i,j\in\cS_U$ with \eqref{U_user}.
			\WHILE{$\phi\left(\alpha^{n,m-1}\right)-f\left(\bP^{n}\right)\geq c\times g_{\bP^n}^{\cM}\Big(\mathrm{grad}_{\cM}f\left(\bP^n\right),$ $\alpha^{n,m-1}\boldsymbol{\eta}^n\Big)$}
			\STATE  Set $\alpha^{n,m}\leftarrow r\alpha^{n,m-1}$ with	  $\alpha^{n,0} = \alpha^0$.
			\STATE Get $\bP^{n,m+1}$ with \eqref{retraction_in} and $\bV_{i,j}^{n,m}, i,j\in\cS_U$ with \eqref{V_in}.
			\STATE Get $\phi\left(\alpha^{n,m}\right)$ with \eqref{phi_nm},   $m\leftarrow m+1$.
			\ENDWHILE
			\STATE  Set $\bP^{n+1}\leftarrow\bP^{n,m}$, $n\leftarrow n+1$.
			\UNTIL{convergence}
		\end{algorithmic}  
	\end{algorithm} 
	\subsection{Computational Complexity}
	
	\renewcommand\arraystretch{2}
	\begin{table}[b]
		\centering
		\caption{The Computational Complexities of Riemannian Ingredients}
		\label{Comp_Riemannian_ingredients}
		\begin{tabular}{cc}
			\bottomrule
			Riemannian ingredients & Computational complexity\\
			\hline
			$\bPi^{T_{\bP}\cN}_{T_{\bP}\cM}\left(\bxi_{\bP}\right)$ & $\sum_{k\in\cS_B}\sum_{i\in\cU_k}M_td_i$ \\
			
			$R_{\bP}^{\cM}\left(\bxi_{\bP}\right)$ & $\sum_{k\in\cS_B}\sum_{i\in\cU_k}M_td_i$  \\ 
			
			$g_{\bP}^\cM\left(\bxi_{\bP},\bzeta_{\bP}\right)$ & $\sum_{k\in\cS_B}\sum_{i\in\cU_k}M_td_i$ \\
			
			$\mathrm{grad}_{\cM}f\left(\bP\right)$ &$O\left(U\sum_{k\in\cS_B}\sum_{i\in\cU_{k}}M_tM_rd_i\right)$\\
			
			$\cT_{\boldsymbol{\eta}_{\bP}}^{\cM}\left(\bxi_{\bP}\right)$&$\sum_{k\in\cS_B}\sum_{i\in\cU_k}M_td_i$ \\
			
			$\bU_{i,j,\ell}, \ell\in\cB_j,i,j\in\cS_U$ &$U\sum_{k\in\cS_B}\sum_{i\in\cU_{k}}M_tM_rd_i$\\
			\toprule
		\end{tabular}
	\end{table}
	
	
	\alref{RCG} is an iterative algorithm and exhibits a fast convergence speed \cite{sato}, where the outer iteration is for obtaining the search direction and the inner iteration is for searching for the step length.  For the $n$-th outer iteration, $\bV_{i,j,\ell}^n,  i,j\in\cS_{U},\ell\in\cB_j,$ defined in \eqref{V_single} can be obtained directly from $\bV_{i,j,\ell}^{n-1}$ and $\bU_{i,j,\ell}^{n-1}$, which have been computed in the $\left(n-1\right)$-th iteration. With $\bV_{i,j,\ell}^{n}$ and \eqref{gradient_Ni_n}, we can get the $\mathrm{grad}_{\cN}f(\bP^n)$, whose computational complexity is $O\left(U\sum_{k\in\cS_B}\sum_{i\in\cU_{k}}M_tM_rd_i\right)$. With the orthogonal projection \eqref{orthogonal_projection}, the Riemannian gradient $\mathrm{grad}_{\cM}f(\bP^n)$ can be derived by projecting $\mathrm{grad}_{\cN}f(\bP^n)$ onto $T_{\bP^n}\cM$ at the cost of $\sum_{k\in\cS_B}\sum_{i\in\cU_k}M_td_i$. Then, the search direction of the current iteration can be obtained from \eqref{eta} by computing $\bU_{i,j,\ell}^{n+1},  \ell\in\cB_j,i,j\in\cS_U$, whose computational complexity is $ U\sum_{k\in\cS_B}\sum_{i\in\cU_{k}}M_tM_rd_i$. 
	
	With the search direction, the step length remains to be determined to reach the next point. During the inner iteration for searching for the step length, the objective function defined in \eqref{phi_nm} needed to be computed and compared for different step lengths to ensure the monotonicity of the proposed method. The objective function can be computed according to \eqref{R_nm}, which is determined by the low dimensional matrix  $\bV_{i,j}^{n,m+1}\in\bbC^{M_r\times d_i}$. Similarly, $\bV_{i,j}^{n,m+1},i,j\in\cS_{U},$ defined in \eqref{V_in} can be obtained directly from $\bV_{i,j,\ell}^n$ and $\bU_{i,j,\ell}^n$, which have been computed before. So we only need to compute the retraction and the $\logdet{\cdot}$ repeatedly during the inner iteration until an efficient $\bP^{n,m+1}$ is reached.  The output of the current iteration is the input of the next iteration. The computational complexities of the elements needed to be computed during an iteration are summarized in \tabref{Comp_Riemannian_ingredients}. We can see that the computational complexities of the orthogonal projection $\bPi^{T_{\bP}\cN}_{T_{\bP}\cM}\left(\bxi_{\bP}\right)$, retraction $R_{\bP}^{\cM}\left(\bxi_{\bP}\right)$, vector transport $\cT_{\boldsymbol{\eta}_{\bP}}^{\cM}\left(\bxi_{\bP}\right)$ and the Riemannian metric $g_{\bP}^\cM\left(\bxi_{\bP},\bzeta_{\bP}\right)$ are the same and much lower than that of the Riemannian gradient and $\bU_{i,j,\ell}, \ell\in\cB_j,i,j\in\cS_U$.

	Let $I_{\mathrm{R}}$, $N_t=BM_t$, $N_r=UM_r$ and $N_d=\sum_{i\in\cS_{U}}d_i$  denote the total numbers of outer iterations of the RCG method,  transmit antennas, receive antennas and data streams, respectively. Let $N_{\mathrm{in}}^n$ denote the number of inner iterations in the $n$-th outer iteration. The computational complexity of the RCG method per inner iteration is particularly low according to the above analyses. Typically, $N_{\mathrm{in}}^n<10$ with $c=10^{-4}$, $r=0.5$ and $\alpha^0=10^{-3}$. Therefore, the computational complexities of the inner iteration during the RCG method can be neglected. The computational complexity of implementing RCG design method on $\cM$ for precoder design in the UCN mMIMO system is $O\left(2U\sum_{k\in\cS_B}\sum_{i\in\cU_{k}}M_tM_rd_i \right)I_{\mathrm{R}}= \frac{\sum_{k\in\cS_{B}}U_k}{BU} O\left( 2N_tN_rN_d \right)I_{\mathrm{R}}\leq O\left( 2N_tN_rN_d \right)I_{\mathrm{R}}$. The popular weighted sum-minimum mean square error (WMMSE) \cite{WMMSE_shi} method has been extended to the  coordinated multi-point joint transmission (CoMP-JT) in the \cite{WMMSE_UCN}, which can be applied in our proposed UCN mMIMO system. The computational complexity of the WMMSE method in the UCN mMIMO system is $O\left( \sum_{k\in\cS_B}\sum_{i\in\cU_{k}}(4UM_tM_rd_i+M_t^2d_i)+BM_t^3 \right)I_{\mathrm{W}}$, where  $I_{\mathrm{R}}$ is the iteration number of the WMMSE method. The computational complexity of the RCG method is much lower than that of the WMMSE method in the case that they have the same number of outer iterations. The computational complexity comparison between the RCG method and several other precoding methods are detailed in \tabref{Comp_Precoders}. Unlike other methods, the computational complexity of the RCG method increases linearly with the total number of transmit antennas $N_t$ and quadratically with the total number of users $U$ in the system.  As the network expands, the enhanced computational efficiency of the RCG method becomes more evident. In the next section, we will present simulation results to further substantiate the superiority of the RCG method proposed.
	
	\renewcommand\arraystretch{2}
	\begin{table}[htbp]
		\centering
		\caption{   The Computational Complexities of Different Precoders  when $\mathrm{B}_{\mathrm{sc}}=3$}
		\label{Comp_Precoders}
		\begin{tabular}{cc}
			\bottomrule
			Precoding Method & Computational complexity\\
			\hline
			RCG & $O\left(\frac{\sum_{k\in\cS_{B}}U_k}{BU}  2N_tN_rN_d \right)I_{\mathrm{R}}$ \\
			
			WMMSE&$O\left( \sum_{k}\!\!\sum_{i\in\cU_{k}}\!(4UM_tM_rd_i\!+\!M_t^2d_i)\!+\!BM_t^3 \right)\!I_{\mathrm{W}}$ \\
			
			ZF & $O(N_r^3+N_tN_r^2)$  \\ 
			
			MMSE & $O(N_t^3+N_t^2N_r)$ \\
			
			BD &$O\left(N_r^2N_tU+N_rN_t^2\right)$\\
			
			EZF &$O\left(N_t^3U+N_rN_t^2\right)$\\
			\toprule
		\end{tabular}
	\end{table}

	\section{Numerical Results}\label{sec_Numerical_Results}
	In this section, we evaluate the performance of the RCG design method in the UCN mMIMO system. We provide extensive simulation results and comparisons under different conditions to validate the superiority of our proposed precoder design and the high computational efficiency of the UCN  mMIMO system.
	\renewcommand\arraystretch{1.8}
	\newcolumntype{"}{@{\hskip\tabcolsep\vrule width 1pt\hskip\tabcolsep}}
	
	\makeatother
	\begin{table}[t]
		\centering
		\caption{Detailed Simulation Parameters}
		\label{Parameter}
		\begin{tabular}{cc"cc}
			\Xhline{1.1pt}
			Center frequency & 4.9 GHz &Speed of each UT &5 km/h\\
			
			Height of each BS & 25 m &Height of each UT&1.5 m \\ 
			
			BS antenna type&3GPP 3D &UT antenna type &ULA \\
			
			$d_i,\forall i\in\cS_U$&2 &$\sigma_z^2$ &-104 dBm \\
			\Xhline{1.1pt}
		\end{tabular}
	\end{table}
	%
	%
	
\begin{figure}[t]
		\centering                 				
		\includegraphics[scale=0.36]{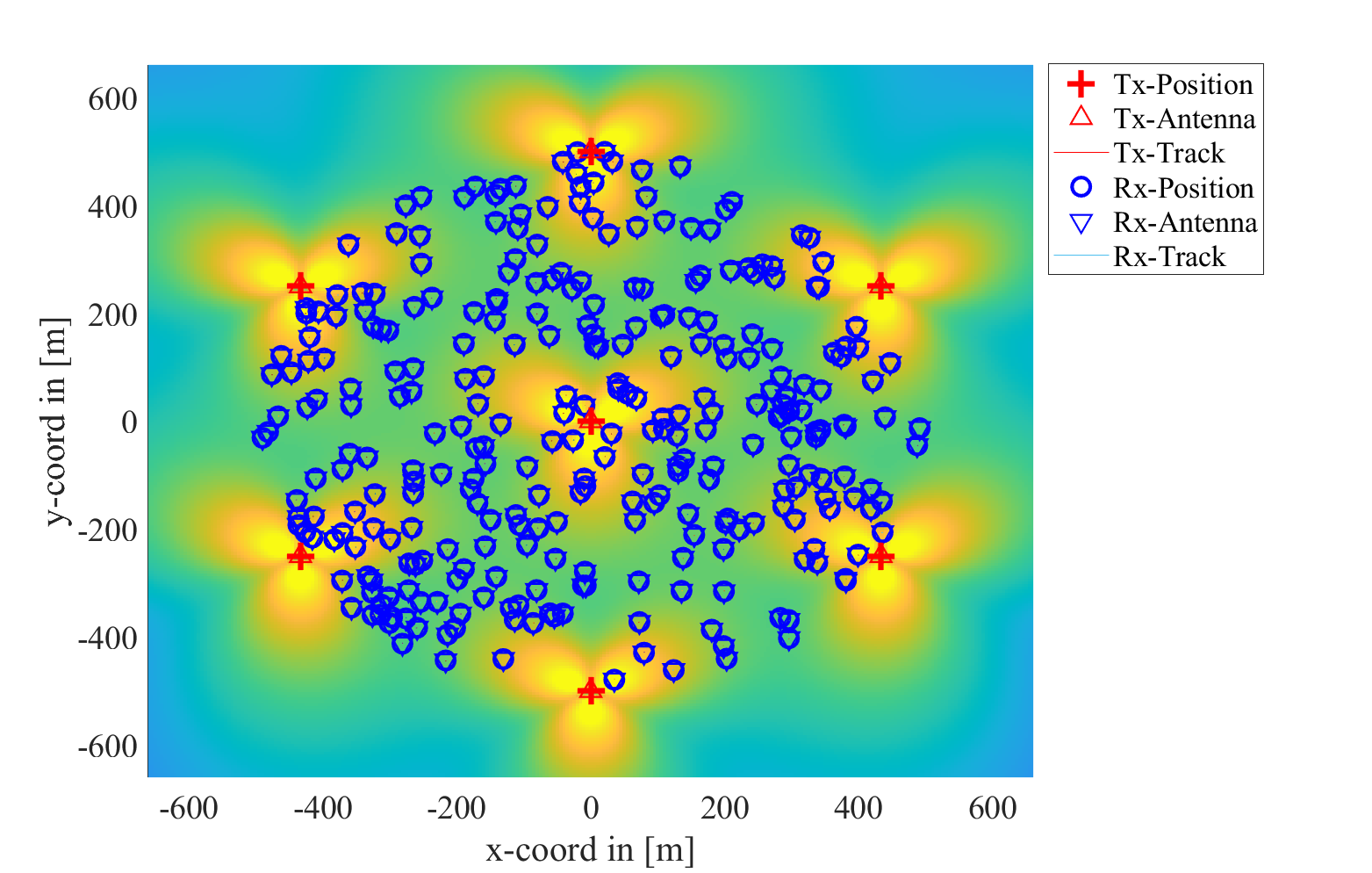}
		\caption{The layout of the UCN mMIMO system.}
		\label{fig_layout}
	\end{figure}

	We adopt  the prevalent QuaDRiGa channel model \cite{QuaDRiGa} to generate a simulation scenario, where ``3GPP 38.901 UMa NLOS” is considered.  To ensure a better coverage, we consider the tri-sector configuration and seven gNodeBs (gNBs) are installed in the system \cite{300}. Each gNB has three co-located BSs and each BS is responsible for a 120-degree coverage \cite{104} as shown in \figref{fig_layout}. So there are totally $B=21$ BSs in the system. The distance between the adjacent  gNBs  is set to 500 m in our simulations \cite{901}.   In the network,  $U=300$ UTs are randomly distributed in a circle with radius of 500 m. For simplicity, we assume $w_1=w_2=\cdots=w_U=1$ and $P_k=P, \forall k$ with $U_k\neq 0$ and $\bP_{\ell}=0, \forall \ell $ with $U_{\ell}=0$, where $P$ is the transmit power that can be adjusted. The serving clusters are formed by selecting the BSs that provide the best channel conditions for each UT \cite{AP}.  Each BS is equipped with $M_t=64$ antennas and each UT has $M_r=2$ antennas. For ease of comparison, we assume that the size of the serving cluster for each UT is the same, specifically denoted as $B_1=B_2=\cdots=B_U=\mathrm{B}_{\mathrm{sc}}$.  More detailed system parameters are summarized in \tabref{Parameter}. For fair comparison, the RCG method and the WMMSE method  are both initialized by the maximum ratio transmission (MRT) \cite{mrt}, which avoids the inverses of large dimensional matrices. It is worth emphasizing that  there is no inverse of large dimensional matrix in our proposed RCG design method with MRT for the initialization.
	
	
	\begin{figure}[t]
		\centering                 				
		\includegraphics[scale=0.5]{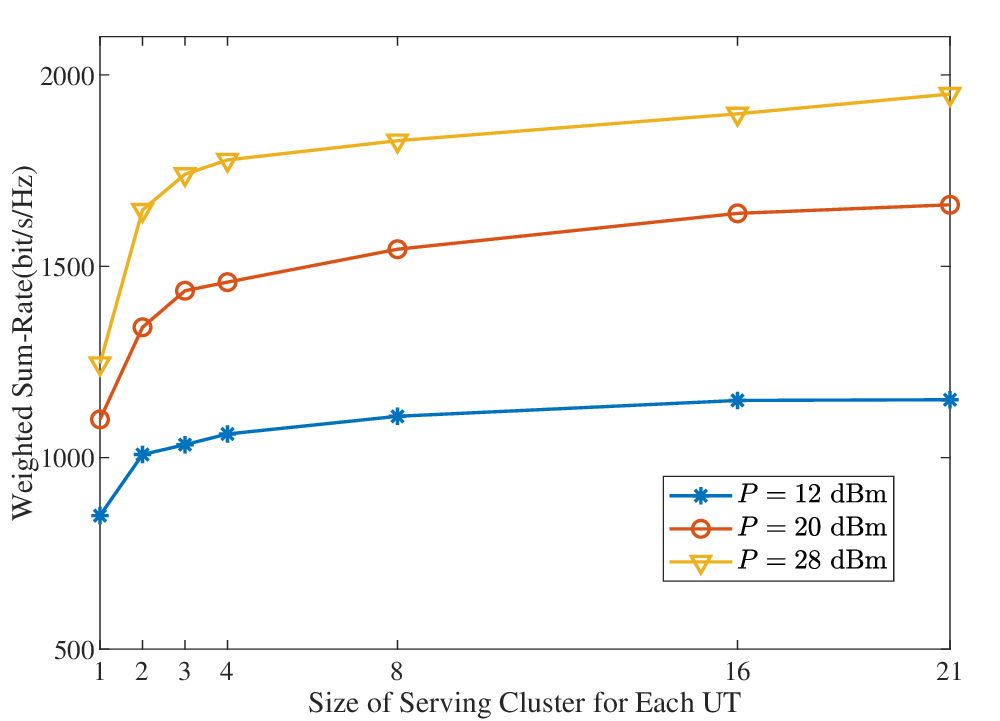}
		\caption{The relationship between the WSR performance and the size of the serving cluster.}
		\label{fig_rateAP}
	\end{figure}

	
	First of all, we study the relationship between the WSR performance and the size of the serving cluster for each UT at different transmit powers in \figref{fig_rateAP}.  As shown in \figref{fig_rateAP}, the WSR performance exhibits a decreasing rate of growth as the size of the serving cluster increases. This is because the BSs having the potential to provide greater service for each UT have been included in the serving cluster. The later the BS is selected into the serving cluster for each UT, the fewer contributions it will make to the WSR performance, and the more interference it could cause to its served group. 
	At $P=28$ dBm, $P=20$ dBm and $P=12$ dBm, it is observed that in the UCN mMIMO system with $\mathrm{B}_{\mathrm{sc}}=21$, only 12\%, 15\% and 11\% WSR performance gains can be achieved, respectively, at the cost of a sevenfold increase in computational complexities compared with the case with $\mathrm{B}_{\mathrm{sc}}=3$. Therefore, the UCN mMIMO system with $\mathrm{B}_{\mathrm{sc}}=3$ can achieve  most of the WSR performance compared with the system with $\mathrm{B}_{\mathrm{sc}}=21$. 
	
	
	
\begin{figure}[t]
		\centering                 				
		\includegraphics[scale=0.5]{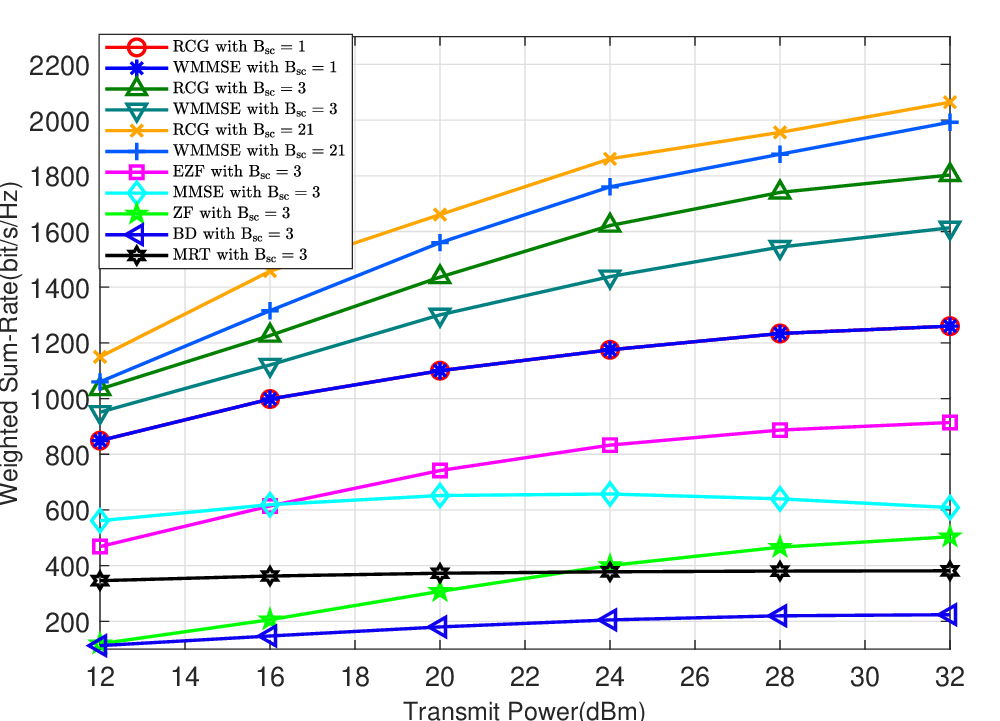}
		\caption{The comparison of the WSR performance between the RCG method and different precoding methods.}
		\label{fig_rate}
	\end{figure}

\begin{figure}[t]
		\centering                 				
		\includegraphics[scale=0.5]{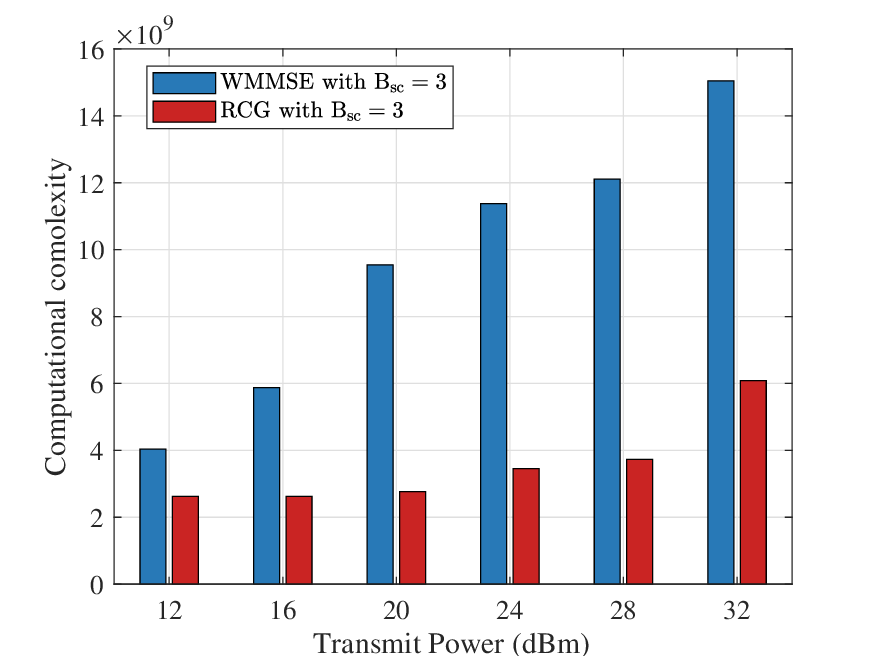}
		\caption{The complexities of the RCG method and the  WMMSE method with $\mathrm{B}_{\mathrm{sc}}=3$ when they have the same performance.}
		\label{fig_barWMMSE}
	\end{figure}

Then, we compare the WSR performance of the RCG method with different precoding methods at different transmit powers in \figref{fig_rate}. Note that the eigen-zero-forcing (EZF), MMSE, ZF, block diagonalization (BD) and MRT precoders are forced to satisfy the power constraints.  From \figref{fig_rate}, we see that the RCG method with $\mathrm{B}_{\mathrm{sc}}=1$ has the same performance as the popular WMMSE method with $\mathrm{B}_{\mathrm{sc}}=1$ \cite{WMMSE_shi}. While in the UCN mMIMO systems with $\mathrm{B}_{\mathrm{sc}}=3$ and $\mathrm{B}_{\mathrm{sc}}=21$, the RCG method all outperforms the WMMSE method \cite{WMMSE_UCN} in the whole transmit power regime. It is worth noting that the RCG method has a much lower computational complexity than the WMMSE method per iteration for a given system, which shows the high efficiency of the RCG method. 
	Although the EZF, MMSE, ZF, BD and MRT methods offer closed-form solutions, their rate performance degrades substantially relative to the RCG method.
  In addition, we observe that the RCG method with $\mathrm{B}_{\mathrm{sc}}=3$ performs much better than the case with $\mathrm{B}_{\mathrm{sc}}=1$ and has a 38\% performance gain when $P=24$ dBm. Although the RCG method with $\mathrm{B}_{\mathrm{sc}}=21$ has the best performance, it suffers from the computational complexity that is seven times higher than that of the RCG method with $\mathrm{B}_{\mathrm{sc}}=3$. To show the high efficiency of the RCG method, we compare the complexities of the RCG method and the WMMSE method in the UCN system with $\mathrm{B}_{\mathrm{sc}}=3$ when the RCG method achieves the same performance as the converged WMMSE method. From \figref{fig_barWMMSE}, we observe that the RCG method needs to pay a much lower computational cost to achieve the same WSR performance as the WMMSE method that have converged.  In fact, the RCG method needs 305.78 s to converge while the WMMSE method requires 1592.31 s when $\mathrm{B}_{\mathrm{sc}}=3$ and $P=24$ dBm, showing the computational efficiency of the RCG method. Note that the simulations are conducted
  	by Matlab R2019b on a desktop with Intel(R) Core(TM) i9-10900K running at 3.70 GHz.
    In addition, the RCG method avoids the inverses of large dimensional matrices and is more advantageous for the forthcoming 6G networks with more antennas equipped at the BS side.
	
	\begin{figure}[t]
		\centering                 				
		\includegraphics[scale=0.5]{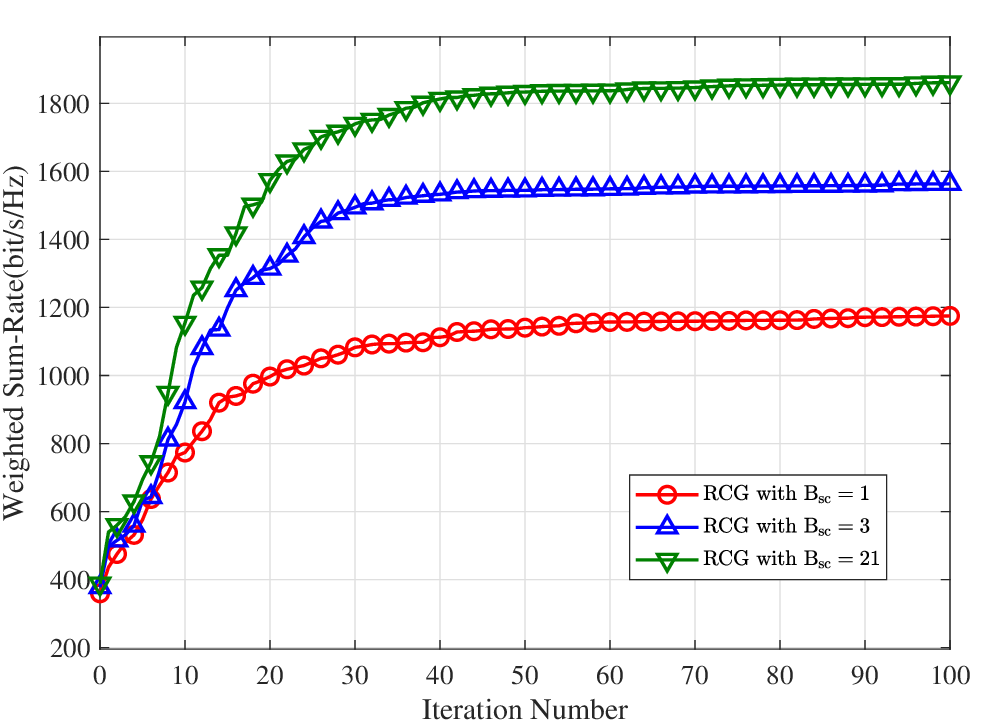}
		\caption{Convergence rate comparison between different configurations when $P=24$ dBm.}
		\label{fig_con_24dbm}
	\end{figure}

	\begin{figure}[t]
		\centering                 				
		\includegraphics[scale=0.5]{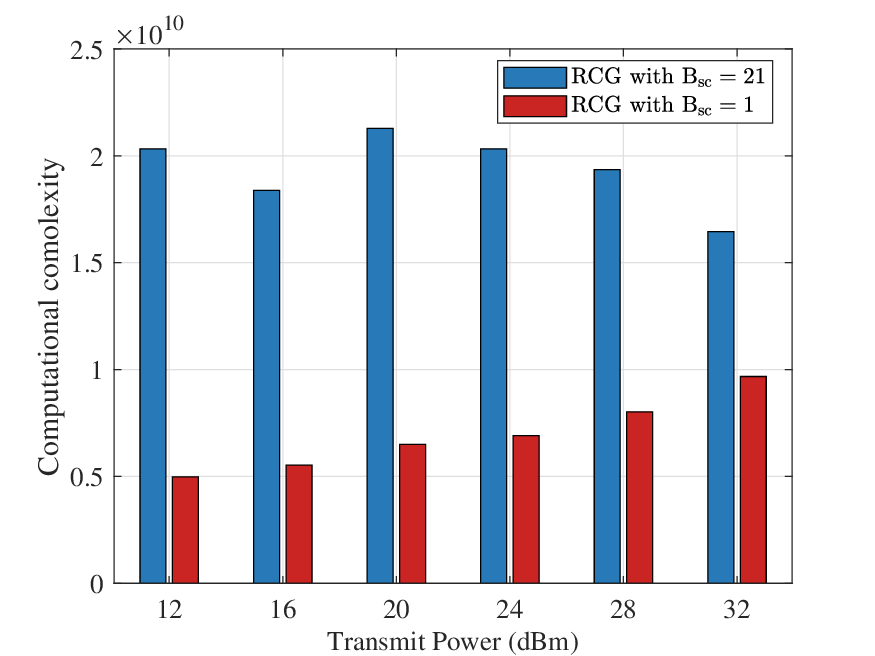}
		\caption{The complexities of the system with $\mathrm{B}_{\mathrm{sc}}=3$ and the  system with $\mathrm{B}_{\mathrm{sc}}=21$ when they have the same performance.}
		\label{fig_bar}
	\end{figure}

	We then study the convergence behavior of our proposed RCG method for precoder design of the UCN mMIMO. In \figref{fig_con_24dbm}, we plot the convergence trajectories of the RCG method when $P=24$ dBm with $\mathrm{B}_{\mathrm{sc}}$ taking different values.  By observing \figref{fig_con_24dbm}, we can see that our method with $\mathrm{B}_{\mathrm{sc}}=3$  has achieved 85\% WSR performance in the first 20 iterations and  93\% performance in the first 30 iterations. 
	The RCG method with $\mathrm{B}_{\mathrm{sc}}=3$ needs about $N=50$ iterations to converge, whose complexity is nearly the same as the RCG method with $\mathrm{B}_{\mathrm{sc}}=21$ when the number of iterations $N=7$. Although the two cases share nearly the same computational complexity, we can see from \figref{fig_con_24dbm} that the system with $\mathrm{B}_{\mathrm{sc}}=3$ and $N=50$ has a large WSR performance gain compared with the system with $\mathrm{B}_{\mathrm{sc}}=21$ and $N=7$.
	Then, we compare the computational complexities of the RCG method when the system with $\mathrm{B}_{\mathrm{sc}}=21$ achieves the same performance as the converged system with $\mathrm{B}_{\mathrm{sc}}=3$  in \figref{fig_bar}. We see that the RCG method with $\mathrm{B}_{\mathrm{sc}}=3$ pays a lower computational cost to achieve the same performance as the case with $\mathrm{B}_{\mathrm{sc}}=21$.
	
	
	\begin{figure}[t]
		\centering                 				
		\includegraphics[scale=0.5]{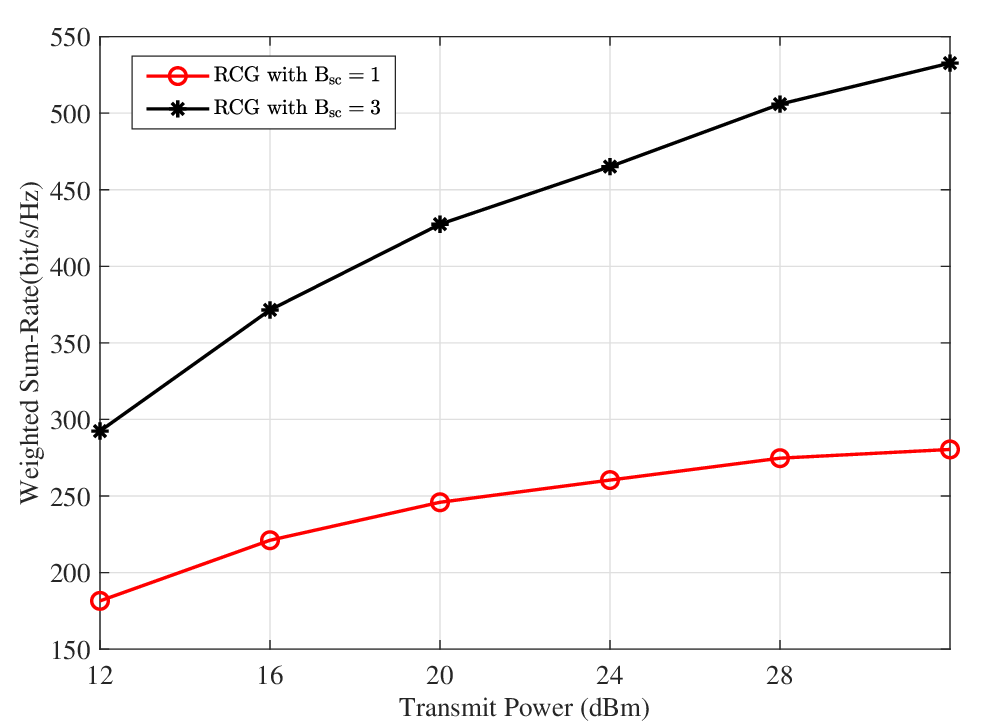}
		\caption{WSR performance of the cell-edge UTs.}
		\label{fig_celledge}
	\end{figure}
	
	To show the performance enhancement of the cell-edge UTs in the UCN mMIMO system, we compare the WSR performance of the cell-edge UTs in the system with $\mathrm{B}_{\mathrm{sc}}=1$ with the system with $\mathrm{B}_{\mathrm{sc}}=3$ in \figref{fig_celledge}. The cell-edge UTs are defined as the UTs suffering higher interference from the adjacent cells according to \cite{celledge_selection}. From \figref{fig_celledge}, we see that the system with $\mathrm{B}_{\mathrm{sc}}=3$ provides a much better WSR performance for cell-edge UTs than the  system with $\mathrm{B}_{\mathrm{sc}}=1$ in the whole transmit power regime. Specifically, the cell-edge UTs have a 79\% performance gain in the system with $\mathrm{B}_{\mathrm{sc}}=3$ compared with the system with $\mathrm{B}_{\mathrm{sc}}=1$ when $P=24$ dBm, showing the superiority of the UCN system with $\mathrm{B}_{\mathrm{sc}}=3$ in enhancing the WSR performance of the cell-edge UTs.
	
	As a matter of fact, the UCN mMIMO system with $\mathrm{B}_{\mathrm{sc}}=1$ can be viewed as the cellular mMIMO system, and the  UCN mMIMO system with $\mathrm{B}_{\mathrm{sc}}=21$ is equivalent to the conventional network mMIMO system. From the above simulation results, it can be inferred that  $\mathrm{B}_{\mathrm{sc}}=3$ represents a favorable choice for the UCN mMIMO system to provide a good enough WSR performance with much reduced computational complexities. Compared with the cellular mMIMO system, the UCN mMIMO system with $\mathrm{B}_{\mathrm{sc}}=3$ can significantly enhance the WSR performance of the UTs in the system, especially for the cell-edge UTs. The computational complexity of the RCG method in the  UCN mMIMO system with $\mathrm{B}_{\mathrm{sc}}=3$ is one-seventh of that in the conventional network system per outer iteration while exhibiting minimal performance degradation.  Additionally, our proposed RCG method for precoder design obviates the need for inverting high-dimensional matrices, and exhibits a better WSR performance and a lower computational complexity compared with the WMMSE method. These results demonstrate  the computational efficiency of the UCN mMIMO system with $\mathrm{B}_{\mathrm{sc}}=3$ and the numerical superiority of our proposed RCG method for precoder design.


	%
		%
		%

	\section{Conclusion}\label{Conclusion}
	In this paper, we have investigated the WSR-maximization precoder design for UCN mMIMO systems with matrix   manifold optimization. In the UCN mMIMO system, the implementation cost of the system and the dimension of the precoder to be designed are much lower than those in the conventional network mMIMO system. By showing the precoders satisfying the power constraints of each BS are on a Riemannian submanifold, we transform the constrained WSR-maximization precoder design problem in Euclidean space to an unconstrained one on the Riemannian submanifold. By deriving all the Riemannian ingredients of the problem on the Riemannian submanifold, the RCG precoder design is proposed for solving the unconstrained problem. The proposed method does not involve the inverses of large dimensional matrices. In addition, the complexity analysis demonstrates the high efficiency of the proposed method. The numerical results not only confirm the superiority of the UCN mMIMO system, but also show significant performance gains and the high computational efficiency of the proposed RCG method for precoder design over the existing methods.

	\appendices
	\section{Proof for \thref{theorem_submanifold}}\label{app_submanifold}
	First, we show that $\cM$ is an embedded submanifold of $\cN$.
	Consider the differentiable function $ F  : \cN\rightarrow \bbR^B : \bP \mapsto F\left(\bP\right) $, and \eqref{M_definition} implies $  \cM = F^{-1}\left( \boldsymbol{0}_B\right)$, where $ \boldsymbol{0}_B \in \bbR^B $. Based on the submersion theorem \cite[Proposition 3.3.3]{Absil2009}, to show $\cM$ is an embedded submanifold
	of $\cN$, we need to prove $ F $ as a submersion at each point of $\cM  $. In other words, we should verify that the rank of $ F $ is equal to the dimension of $ \bbR^B $, i.e., $B$, at every point of $\cM$. Let  $\bZ =  \bZ_1\times\bZ_2\times\cdots\times\bZ_U   $  be an arbitrary point on $\cN$.  Since the rank of $ F $ at $ \bP \in \cN $ is defined as the dimension of the range of $ \mathrm{D}F\left( \bP\right)$, we need to show that for all $ \bw\in\bbR^B $,  there exists $ \bZ \in \cN $  such that $ \mathrm{D}F\left( \bP\right)\left[ \bZ\right] = \bw^T=\left(w_1,w_2,\cdots,w_B\right) $. Since the differential operation at  $ \bP $ is equivalent to the component-wise differential at  each of $ \bP_{1},\bP_{2},\cdots, \bP_U$, we have 
	\begin{equation}\label{key}
		\begin{aligned}
			\mathrm{D}F\left( \bP\right)\left[ \bZ\right] =& \Big( \sum_{k\in\cB_1}2\Re\left\{\mathrm{tr}\left(\bP_1^H\bW_1^H\bQ_k\bW_1\bZ_1\right)\right\},\cdots,\\
			&\sum_{k\in\cB_U}2\Re\left\{\mathrm{tr}\left(\bP_U^H\bW_U^H\bQ_k\bW_U\bZ_U\right)\right\}\Big).
		\end{aligned}
	\end{equation}
	By choosing $ \bZ_i = \frac{1}{2} \frac{w_k}{P_k} \bP_i$, we will have $\mathrm{D}F\left( \bP\right)\left[ \bZ\right] = \bw^T$. This shows that $ F $ is full rank as well as a submersion on $\cM$, and $ \cM $ is an embedded submanifold of $\cN $.

	In this case, $T_{\bP}{\cM}$ can be regarded as a subspace of $T_{\bP}{\cN}$, and the Riemannian metric $g_{\bP}^{\cN}\left(\cdot\right)$ on $\cN$ naturally induces a Riemannian metric $g_{\bP}^{\cM}\left(\cdot\right)$ on $\cM$ according to
	\begin{equation}\label{RiemannianMetricSubmanifold}
		\begin{aligned}
			g_{\bP}^{\cM}\left(\bxi_{\bP},\bzeta_{\bP}\right)=g_{\bP}^{\cN}\left(\bxi_{\bP},\bzeta_{\bP}\right),
		\end{aligned}
	\end{equation}
	where $\bxi_{\bP}\in T_{\bP}\cM$ and $\bzeta_{\bP}\in T_{\bP}\cM$ on the right hand side are viewed as elements in $T_{\bP}\cN$.  With this metric, $ \cM  $ is a Riemannian submanifold of $ \cN $.
	
	\section{Proof for \lmref{lemma_projection}}\label{app_projection}
	With \eqref{decomposition_xi} and \eqref{normal_space_M}, we have
	\begin{equation}
		\begin{aligned}
			&\bPi^{T_{\bP}\cN}_{T_{\bP}\cM}\left(\bxi_{\bP}\right)=\Big( \bxi_{\bP_1}-\sum_{\ell\in\cB_1}\mu_\ell \bW_1^H\bQ_\ell\bW_1\bP_1,\cdots,\\
			&\bxi_{\bP_U}-\sum_{\ell\in\cB_U}\mu_\ell \bW_U^H\bQ_\ell\bW_U\bP_U \Big)\in T_{\bP}\cM,
		\end{aligned}
	\end{equation}
	which satisfies \eqref{tangent_space_M}. So the equation
	\begin{equation}
		\begin{aligned}
			\sum_{i\in\cU_{k}}\Re\Big\{&\mathrm{tr}\Big(  \big(\bxi_{\bP_i}-\sum_{\ell\in\cB_i}\mu_{\ell}\bW_i^H\bQ_{\ell}\bW_i\bP_i\big)^H\\
			&\times\bW_i^H\bQ_k\bW_i\bP_i\Big)\Big\}=0
		\end{aligned}
	\end{equation}
	holds for $ k\in\cS_{B}$. After some algebra, for $\ell\in\cS_{B}$, we can get
	\begin{equation}
\begin{aligned}
	\mu_\ell=\frac{1}{P_{\ell}}\sum_{i\in\cU_{\ell}}\Re\left\{ \mathrm{tr}\left( \bP_i^H\bW_i^H\bQ_\ell\bW_i
	\bxi_{\bP_i} \right) \right\}.
\end{aligned}
	\end{equation}
	\section{Proof for \thref{theo_gradient}}\label{app_gradient}
	Given that $\cN$ is the product linear manifold composed of $\sum_{i\in\cS_{U}}B_i$ complex vector spaces, $\mathrm{grad}_{\cN}f\left(\bP\right)$ only depends on $f\left(\bP\right)$. For notational simplicity, we use $f\left(\bP_i\right)$ to denote the objective function that only considers $\bP_i$ as the optimization variable with $\bP_j, \forall j\neq i$, fixed. Similarly, let $f\left(\bP_{i,k}\right)$ denote the objective function that only considers $\bP_{i,k},  k\in\cB_i,$ as the optimization variable with $\bP_{j,\ell}, \forall \left(j,\ell\right)\neq \left(i,k\right)$, fixed. $\mathrm{grad}_{\cN}f\left(\bP\right)$ is made up of $\mathrm{grad}_{\cN_{i}}f\left(\bP_{i}\right),\forall i \in\cS_{U}$ as shown in \eqref{tangent_space_N} and  $\mathrm{grad}_{\cN_{i}}f\left(\bP_{i}\right)$ is made up of $\mathrm{grad}_{\cN_{i,k}}f\left(\bP_{i,k}\right),\forall k \in\cB_i,$ as shown in \eqref{tangent_space_Ni}. So we derive $\mathrm{grad}_{\cN_{i,k}}f\left(\bP_{i,k}\right)$ first. For any $\bxi_{\bP_{i,k}}\in T_{\bP_{i,k}}\cN_{i,k}$, the directional derivative of $  f \left( \bP_{i,k}\right)  $ along $ \bxi_{\bP_{i,k}} $ is  
	\begin{equation}\label{Df_Pik}
		\begin{aligned}
			&\mathrm{D} f \left( \bP_{i,k}\right)\left[ \bxi_{\bP_{i,k}}\right]   =\\
			& -w_i \mathrm{D}\cR_{{i,k}}^i\left[ \bxi_{\bP_{i,k}}\right] - \sum_{ j \neq i}^U w_{j} \mathrm{D}\cR_{i,k}^{j}\left[ \bxi_{\bP_{i,k}}\right],  i\in\cS_{U}, k\in\cS_{B},
		\end{aligned}
	\end{equation}
	where $\cR_{{i,k}}^i$ is the rate of UT $i$ that only considers $\bP_{i,k}$ as the optimization variable and $\cR_{i,k}^{{j}}$ is the rate of UT $j$ that only considers $\bP_{i,k}$ as the optimization variable. $\cR_{{i,k}}^i$ and $\cR_{i,k}^{{j}}$ can be  viewed as mappings from $\bbC^{M_t\times d_i}$ to $\bbR$, so $ \mathrm{D}\cR_{i,k}^i\left[ \bxi_{\bP_{i,k}}\right] $ and $  \mathrm{D}\cR_{i,k}^{j}\left[ \bxi_{\bP_{i,k}}\right] $ can be obtained from \eqref{DFX}.
	We derive $ \mathrm{D}\cR_{i,k}^i\left[ \bxi_{\bP_{i,k}}\right] $ and $  \mathrm{D}\cR_{i,k}^{j}\left[ \bxi_{\bP_{i,k}}\right] $ separately as 
	\begin{equation}
		\begin{aligned}
			\mathrm{D} \cR_{i,k}^i \left[ \bxi_{\bP_{i,k}}\right]  =&  \mathrm{tr}\bigg(\bC_{i}  \Big( \bxi_{\bP_{i,k}}^H\bH_{i,k}^H\bR_i^{-1}\bH_{i}\bW_i\bP_{i}+\\
			&\bP_{i}^H\bW_i^H\bH_{i}^H\bR_i^{-1}\bH_{i,k}\bxi_{\bP_{i,k}}\Big)\bigg) \\
			=& g_{\bP_{i,k}}^{\cN_{i,k}} \left(2 \bH_{i,k}^H \bR_i^{-1}\bH_{i}\bP_{i}\bC_i, \bxi_{\bP_{i,k}}\right),
		\end{aligned}
	\end{equation} 
	\begin{equation}
		\begin{aligned}
			\mathrm{D}\cR_{i,k}^{j}\left[ \bxi_{\bP_{i,k}}\right] 
			&= -\mathrm{tr}\bigg(\bH_{j,k}^H\bR_{j}^{-1}\bH_{j}\bW_j\bP_{j}\bC_{j}\bP_{j}^H\bW_j^H\bH_{j}^H \\
			&\times \bR_{j}^{-1}\bH_{j}\left( \bxi_{\bP_{i,k}}\bP_{i}^H\bW_i^H + \bW_i\bP_{i} \bxi_{\bP_{i,k}}^H \right)  \bigg)   \\
			=& g_{\bP_{i,k}}^{\cN_{i,k}} \Big( -2\bH_{j,k}^H\bR_{j}^{-1}\bH_{j}\bW_j\bP_{j}\bC_{j}\\
			&\times \bP_{j}^H\bW_j^H\bH_{j}^H\bR_{j}^{-1}\bH_{j}\bW_i\bP_{i}, \bxi_{\bP_{i,k}}    \Big).
		\end{aligned}
	\end{equation}
	Thus, we have 
	\begin{equation}
		\begin{aligned}
			&\mathrm{D} f\left( \bP_{i,k}\right)\left[ \bxi_{\bP_{i,k}}\right] 
			= g_{\bP_{i,k}}^{\cN_{i,k}}\big(\bxi_{\bP_i},-2  w_i\bH_{i,k}^H \bR_i^{-1}\bH_{i}\bW_i\bP_{i}\bC_i \\
			+&2\sum_{ j \neq i}w_{j}\bH_{j,k}^H\bR_{j}^{-1}\bH_{j}\bW_j\bP_{j}\bC_{j}\bP_{j}^H\bW_j^H\bH_{j}^H\bR_{j}^{-1}\bH_{j}\bW_i\bP_{i}  \big)
		\end{aligned}
	\end{equation}
	and $ \mathrm{grad}_{\cN_{i,k}} f \left( \bP_{i,k}\right) $ is 
	\begin{equation}\label{gradient_Nik}
		\begin{aligned}		
			&\mathrm{grad}_{\cN_{i,k}} f \left( \bP_{i,k}\right) = -2\big( w_i\bH_{i,k}^H \bR_i^{-1}\bH_{i}\bW_i\bP_{i}\bC_i \\
			&- \sum_{j\neq i}w_{j}\bH_{j,k}^H\bR_{j}^{-1}\bH_{j}\bW_j\bP_{j}\bC_{j}\bP_{j}^H\bW_j^H\bH_{j}^H\bR_{j}^{-1}\bH_{j}\bW_i\bP_{i}\big).
		\end{aligned}
	\end{equation} 
	With  \eqref{tangent_space_Ni}, \eqref{tangent_space_N} and \eqref{gradient_Nik}, $\mathrm{grad}_{\cN}f\left(\bP\right)$ can be easily obtained.
	With \cite[Theorem 3.6.1]{Absil2009} and \lmref{lemma_projection}, the Riemannian gradient of $ f \left(\bP\right)  $ in $T_{\bP}\cM$ is 
	\begin{equation}\label{gradient_sub}
		\begin{aligned}
			\mathrm{grad}_{\cM}& f \left( \bP \right)  = \bPi_{T_{\bP}\cM}^{T_{\bP}\cN} \left( \mathrm{grad}_{\cN} f\left( \bP \right) \right) \\
			=\Big(&\mathrm{grad}_{\cN_1}f\left(\bP_1\right)-\sum_{k\in\cB_1}\lambda_k\bW_1^H\bQ_k\bW_1\bP_1,\cdots,\\
			&\mathrm{grad}_{\cN_U}f\left(\bP_U\right)-\sum_{k\in\cB_U}\lambda_k\bW_U^H\bQ_k\bW_U\bP_U\Big),
		\end{aligned}
	\end{equation}
	where 
	\begin{equation}
		\begin{aligned}
			\lambda_k=\frac{1}{P_{k}}\sum_{i\in\cU_{k}}\Re\left\{ \mathrm{tr}\left( \bP_i^H\bW_i^H\bQ_k\bW_i
			\mathrm{grad}_{\cN_i}f\left(\bP_i\right) \right) \right\}.
		\end{aligned}
	\end{equation}
	\section{Proof for \thref{theo_vector_transport}}\label{app_vector_transport}
	From \cite[Section 8.1.3]{Absil2009}, the vector transport for a Riemannian submanifold of  a Euclidean space can be defined by the orthogonal projection, based on which the vector transport is the orthogonal projection from $T_{\bP}\cM$ to $T_{\bP^{\mathrm{new}}}\cM$. We have obtained the orthogonal projection from $T_{\bP}\cN$ to $T_{\bP}\cM$ in \lmref{lemma_projection} based on that $T_{\bP}\cN$ can be decomposed into two orthogonal spaces. Given that $T_{\bP}\cM$ is a subspace of $T_{\bP}\cN=\bbC^{B_1M_t\times d_1}\times\cdots\times\bbC^{B_UM_t\times d_U}=T_{\bP^{\mathrm{new}}}\cN$, $T_{\bP}\cM$ can also be decomposed as
	\begin{equation}\label{decoposition_sub}
		\begin{aligned}
			T_{\bP}\cM=T_{\bP^{\mathrm{new}}}^{\mathrm{sub}}\cM\oplus N_{\bP^{\mathrm{new}}}^{\mathrm{sub}}\cM,
		\end{aligned}
	\end{equation}
	where $T_{\bP^{\mathrm{new}}}^{\mathrm{sub}}\cM$ and $N_{\bP^{\mathrm{new}}}^{\mathrm{sub}}\cM$ are the subspaces of the tangent space and the normal space of $\cM$ at $\bP^{\mathrm{new}}$, respectively. From \eqref{normal_space_M}, we have
	\begin{equation}\label{NPM_sub}
		\begin{aligned}
			N_{\bP^{\mathrm{new}}}^{\mathrm{sub}}\cM=&\Big\{ \Big( \sum_{\ell\in\cB_1}\mu_\ell\bW_1^H \bQ_\ell\bW_1\bP_1^{\mathrm{new}},\cdots,\\
			&\sum_{\ell\in\cB_U}\mu_\ell\bW_U^H \bQ_\ell\bW_U\bP_U^{\mathrm{new}} \Big) \mid  \rho_\ell\in\bbR, \bP\in\cM \Big\}.
		\end{aligned}
	\end{equation}
	From \eqref{decoposition_sub} and \eqref{NPM_sub}, the vector transport can be defined as
	\begin{equation}
		\begin{aligned}
			\cT_{\boldsymbol{\eta}_{\bP}}^{\cM}\left(\bxi_{\bP}\right)=&\bPi^{T_{\bP}\cM}_{T_{\bP^{\mathrm{new}}}\cM}\left(\bxi_{\bP}\right)=\bPi^{T_{\bP}\cM}_{T_{\bP^{\mathrm{new}}}^{\mathrm{sub}}\cM}\left(\bxi_{\bP}\right)\\
			=&\Big(\bxi_{\bP_1}-\sum_{\ell\in\cS_B}\rho_{\ell}\bW_1^H\bQ_{\ell}\bW_1\bP^{\mathrm{new}}_1,\cdots,\\
			&\bxi_{\bP_U}-\sum_{\ell\in\cS_B}\rho_{\ell}\bW_U^H\bQ_{\ell}\bW_U\bP^{\mathrm{new}}_U\Big).
		\end{aligned}		
	\end{equation}
	As $T_{\bP^{\mathrm{new}}}^{\mathrm{sub}}\cM\subseteq T_{\bP^{\mathrm{new}}}\cM$, $\left(\bxi_{\bP}-\sum_{\ell\in\cS_B}\rho_\ell\bQ_\ell\bP^{\mathrm{new}}\right)$ should satisfy
	\begin{equation}
		\begin{aligned}
			\sum_{i\in\cU_{k}}\Re&\Big\{\mathrm{tr}\Big(\left(\bP_i^{\mathrm{new}}\right)^H\bW_i^H\bQ_k\bW_i\\
			&\times\big(\bxi_{\bP_i}-\sum_{\ell\in\cS_B}\rho_\ell\bW_i^H\bQ_\ell\bW_i\bP_i^{\mathrm{new}}\big)\Big)\Big\}=0
		\end{aligned}
	\end{equation}
	After some algebra, we can get
	\begin{equation}
		\rho_{\ell}=\frac{1}{P_{\ell}}\sum_{i\in\cU_{\ell}}\Re\left\{\mathrm{tr}\left(\left(\bP^{\mathrm{new}}_i\right)^H\bW_i^H\bQ_\ell\bW_i\bxi_{\bP_i}\right)\right\}, \ell \in \cS_B.
	\end{equation}
	
	\bibliographystyle{IEEEtran}
	\bibliography{Reference}

\end{document}